\documentclass[]{article}
\usepackage{arxiv}

\usepackage{subcaption}
\usepackage{placeins}
\usepackage{mathtools}
\usepackage{amssymb}
\usepackage{amsmath}
\usepackage{amsfonts}   
\usepackage{fancyhdr}
\usepackage{times}
\usepackage{amsthm}
\usepackage{siunitx}
\usepackage{nicefrac}
\usepackage{wrapfig}

\newtheorem{theorem}{Theorem}%

\usepackage{todonotes}
\usepackage{algorithm}
\usepackage{algorithmic}

\usepackage{array}

\newcommand\myarray[1]{%
  \begingroup
  \renewcommand\arraystretch{1.33}
  \left\{ \begin{array}{@{}l@{}} #1 \end{array} \right.
  \endgroup}

\newcommand{\K}{\mathcal{K}}

\newcommand{\diff}{\mathrm{d}}
\newcommand{\given}{\,|\,}

\newcommand{\pit}{\tilde{\pi}}

\newcommand{\rel}{\sim}

\newcommand{\subdivision}{\mathcal{S}}

\newcommand{\defeq}{\overset{\text{def}}{=}}

	\usepackage{tikz}
	\usetikzlibrary{shapes.geometric, calc, positioning, intersections}
	\usepackage{adjustbox}

	\definecolor{color1}{RGB}{34, 168, 132}
	\definecolor{color2}{RGB}{42, 120, 142}
	\definecolor{color3}{RGB}{68, 1, 84}

\usepackage{booktabs}  
\usepackage{nicefrac}
\usepackage{microtype}

\DeclareMathOperator{\Dir}{Dir}

\usepackage{float}

\usepackage{enumitem}

\usepackage{xcolor}
\definecolor{darkblue}{RGB}{0,0,128}
\definecolor{darkgreen}{RGB}{0, 128, 0}
\definecolor{darkred}{RGB}{128, 0, 0}
\definecolor{black}{RGB}{0, 0, 0}
\definecolor{errorcolor}{HTML}{481567}
\definecolor{viridisgreen}{HTML}{55C667}

\usepackage[backend=biber,
style=apa,
maxcitenames=2,
maxbibnames=100, 
language=english,
doi=false,
isbn=false,
url=false,
uniquename=false
]{biblatex} 
\DeclareLanguageMapping{english}{english-apa} %
\usepackage{hyperref}
\usepackage{url}  
\hypersetup{
	colorlinks=true,
	linkcolor=darkblue,
	filecolor=darkblue,
	urlcolor=darkblue,
	citecolor=darkblue,
	pdfauthor={},
	pdftitle={}
}

\title{The Simplex Projection: Lossless Visualization of \\4D Compositional Data on a 2D Canvas}
\author{
    Marvin Schmitt\thanks{Corresponding author: Marvin Schmitt, \href{mailto:mail.marvinschmitt@gmail.com}{mail.marvinschmitt@gmail.com}}\\
    University of Stuttgart
    \And
    Yuga Hikida\\
    TU Dortmund University
    \AND
    Stefan T.\ Radev\\
    Rensselaer Polytechnic Institute
    \And
    Filip Sadlo\\
    Heidelberg University
    \And
    Paul-Christian Bürkner\\
    TU Dortmund University
}

\begin{document}

\maketitle

\begin{abstract}
The simplex projection expands the capabilities of simplex plots (also known as ternary plots) to achieve a lossless visualization of 4D compositional data on a 2D canvas. Previously, this was only possible for 3D compositional data. We demonstrate how our approach can be applied to individual data points, point clouds, and continuous probability density functions on simplices. While we showcase our visualization technique specifically for 4D compositional data, we offer rigorous proofs that support its extension to compositional data of any (finite) dimensionality.
\end{abstract}

\section{Introduction}
The visualization of high-dimensional data is a key task in countless domains of scientific research. 
Yet, the representation of multi-dimensional data in a two-dimensional canvas (e.g., static screens or paper) can pose a significant challenge, leading to substantial information loss or distortions, which, in turn, can skew the interpretation and analysis of the data.

In this paper, we address this challenge by developing a novel approach for visualizing 4D compositional data on a 2D canvas.
Compositional data consists of vectors with strictly positive entries that sum to one \autocite{Greenacre2021}. This data type naturally arises for proportions, normalized data, or discrete probabilities.
Examples for compositional data include (i) the relative composition of the gut microbiome \autocite{Gloor2017}; (ii) proportion of peoples' activities throughout the day \autocite[e.g., activity, rest, and sleep;][]{Dumuid2020}; or (iii) discrete probability vectors, such as posterior model probabilities arising in Bayesian model comparison \autocite{PhilipDawid2011,schmitt2023metauncertainty}.

Our technique, which we call \emph{simplex projection}, is a lossless visualization method that accurately represents the compositional data while preserving its geometrical and topological properties. 
We prove mathematically that our mapping from 4D compositional data to its 2D representation is a bijection (invertible one-to-one correspondence) that incurs no loss of information.
We demonstrate the effectiveness of our approach, highlighting the simplex projection as a potent tool for exploring and analyzing 4D compositional data.
While the underlying mathematical treatment holds for arbitrary finite dimensions, throughout the paper, we will focus chiefly on illustrations and intuitions for the 4D case.

\section{Preliminaries}\label{sec:preliminaries}
Throughout this manuscript, let $J\in\mathbb{N}$, and the points~$v_1, \ldots, v_J\in\mathbb{R}^J$ be affinely independent, that is, $(v_2-v_1), \ldots, (v_j-v_1)$ are linearly independent.
Further, the points~$v_1, \ldots, v_J$ are the vertices of the $(J-1)$-dimensional simplex $\Delta^{J-1}$ defined by the set
\begin{equation}\label{eq:simplex-definition}
    \Delta^{J-1} = 
    \Big\{
        x\in\mathbb{R}^J: x = \sum_{j=1}^J\pi_j v_j
    \Big\},
\end{equation}    
with weights $\pi_j\in(0,1)$ such that $\sum_{j=1}^J\pi_j=1$.
When the dimension of the simplex~$\Delta^{J-1}$ is sufficiently clear from the context, we drop the superscript and simply write $\Delta$.

The representation of a point $x$ through weighted vertices, that is, via coefficients~$(\pi_1,\ldots,\pi_J)$ in \autoref{eq:simplex-definition}, is commonly referred to as \emph{barycentric coordinates} with respect to the vertices~$v_1, \ldots, v_J$.
For brevity, we will slightly overload the notation and use the vector of barycentric coordinates $(\pi_1, \ldots, \pi_J)$ to refer to a point~$x=\sum_{j=1}^J\pi_j\,v_j$ in the simplex with vertices $v_1,\ldots,v_J$ as defined in \autoref{eq:simplex-definition}.
Since any two simplices of equal dimension $J$ are homeomorphic by a simplicial homeomorphism \autocite{lee_introduction_2000}, the exact location of the vertices $v_1, \ldots, v_J$ is irrelevant and we will use regular (aka.\ equilateral) simplices for all illustrations.

The convex hull of each non-empty subset of size $N$ from the $J$ vertices $v_1,\ldots,v_J$ of a simplex $\Delta^{J-1}$ is called a $(N-1)$-face.
In particular, the $0$-faces are the vertices, the $1$-faces are the edges, the $(J-2)$-faces are the facets which we denote as $\sigma$, and the only $(J-1)$-face is the simplex $\Delta^{J-1}$ itself. 
We denote the facet opposing a given vertex~$v_j$ as $\sigma_{-j}$.

\begin{figure}[t]
\begin{minipage}{0.47\linewidth}
    \begin{figure}[H]
        \centering
        \includegraphics[width=\linewidth]{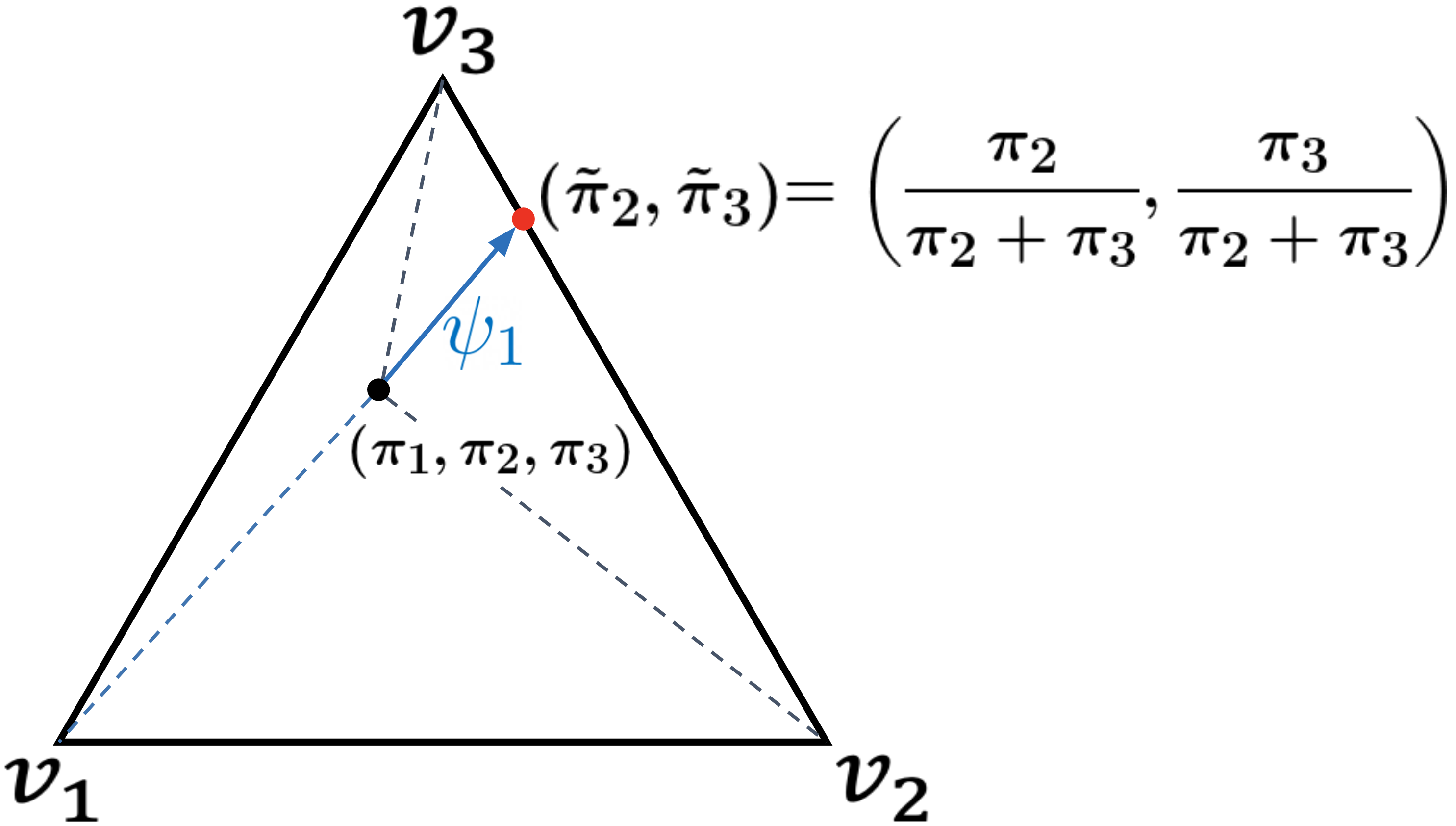}
        \caption{Projection of $x=(\pi_1, \pi_2, \pi_3)$ onto the facet (edge) $\sigma_{-1}=\overline{v_2v_3}$ in a triangle (2-simplex). It can be seen clearly that the perspective projection $\psi_1$ does not change the points' coordinate ratios, $\pi_2/\pi_3 = \tilde{\pi}_2 / \tilde{\pi}_3$.}
        \label{fig:invariance_ratio}
    \end{figure}
\end{minipage}
\hfill %
\begin{minipage}{0.48\linewidth}
    \begin{figure}[H]
        \centering
        \begin{subfigure}[t]{0.49\linewidth}
            \includegraphics[width=\linewidth]{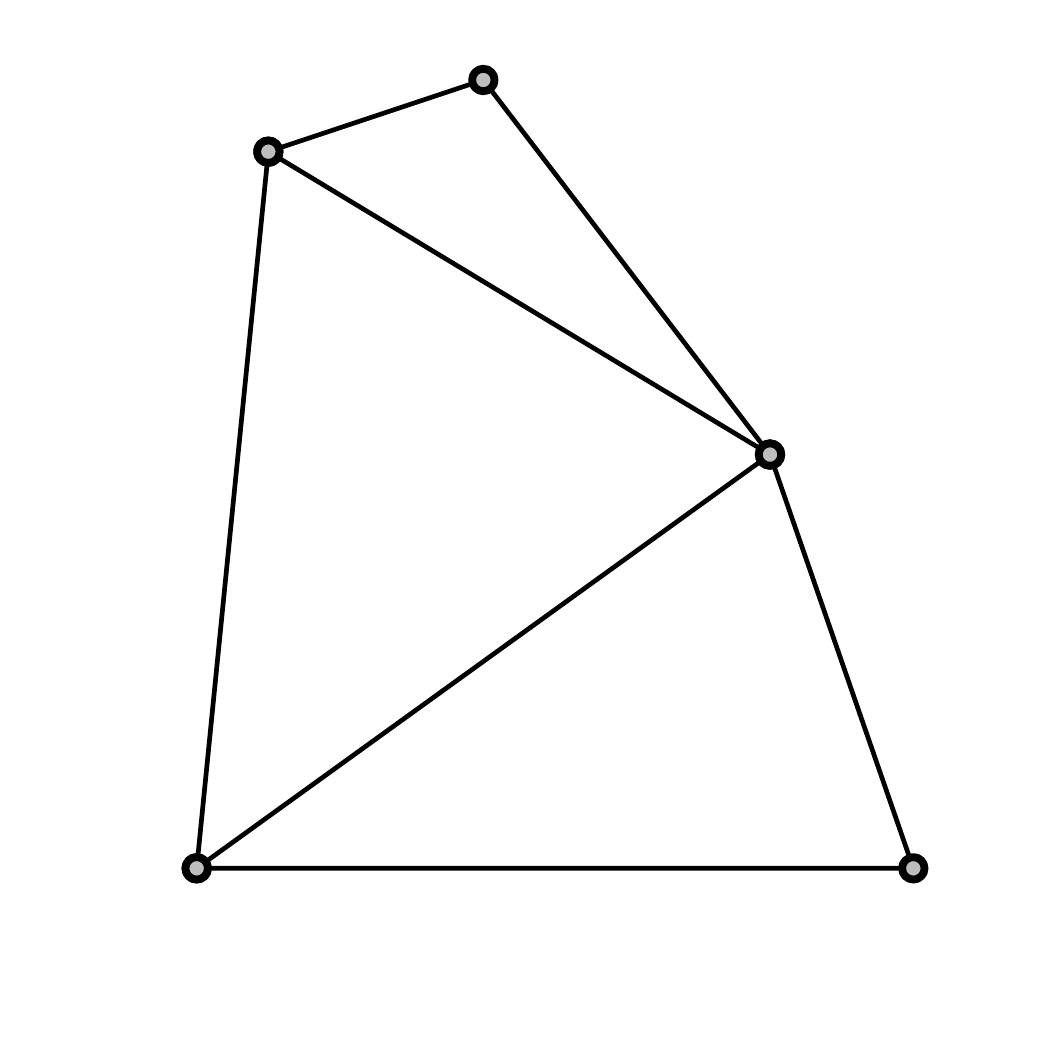}
            \caption{Simplicial complex.}
        \end{subfigure}
        \hfill
        \begin{subfigure}[t]{0.49\linewidth}
            \includegraphics[width=\linewidth]{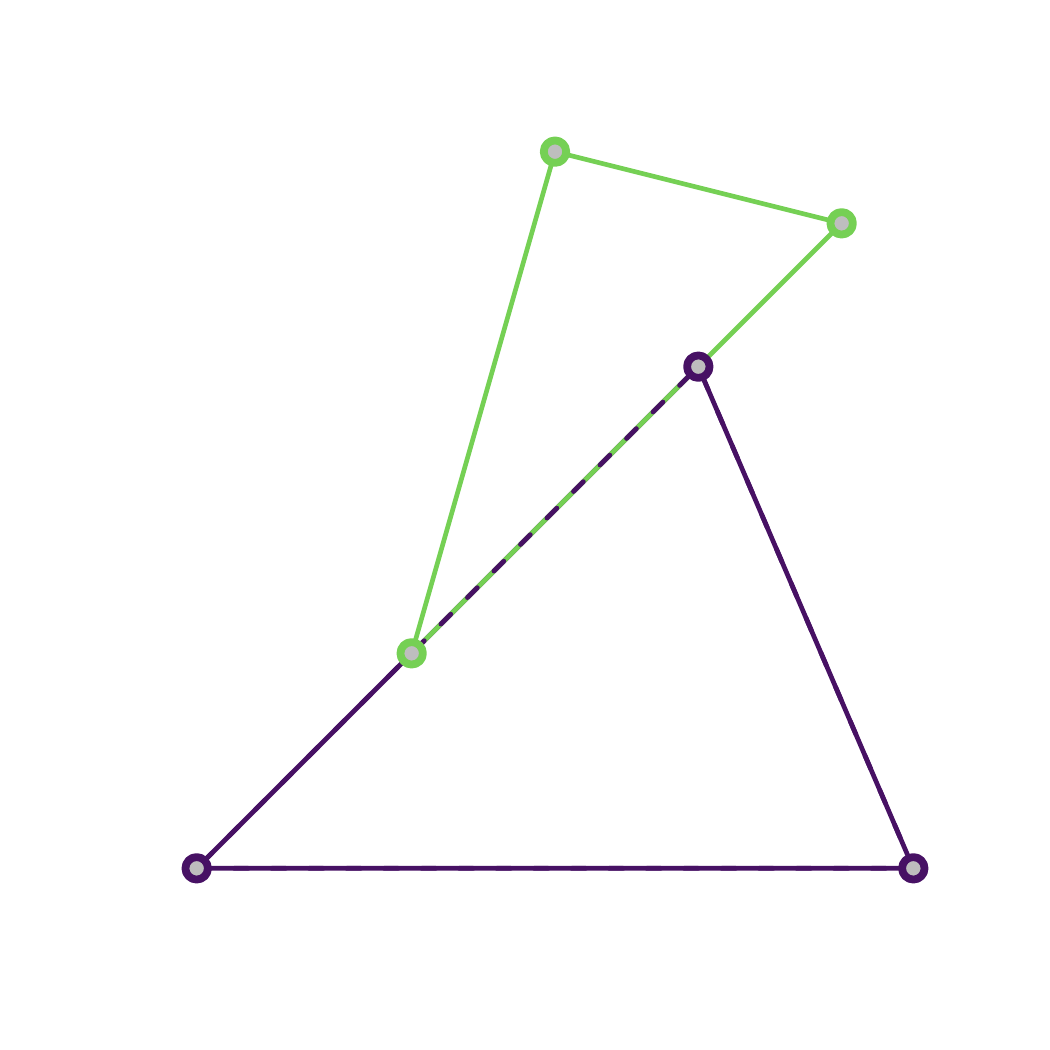}
            \caption{No simplicial complex.}
        \end{subfigure}
        \caption{Not all sets of simplices are simplicial complexes. In (a), three $2$-simplices form a (pure) simplicial complex. In (b), the dashed intersection of the 2-simplices is neither empty nor a face of the two simplices.}
        \label{fig:simplicial_complex}
    \end{figure}
\end{minipage}
\end{figure}

\paragraph{Renormalized Barycentric Coordinates}
Let $(\pi_1, \ldots, \pi_J)$ be barycentric coordinates with $J$ components, as defined above.
For an index subset $K\subseteq \{1,\ldots, J\}$, we define 
\begin{equation}\label{eq:renormalized-barycentric-coordinates}
    \Big(\pit_k\Big)_{k\in K} =
    \left(\dfrac{\pi_k}{\sum\limits_{k{'}\in K}\pi_{k{'}}}\right)_{k\in K}
\end{equation}
as \emph{renormalized barycentric coordinates} (mind the tilde to differentiate between vanilla and renormalized coordinates).
The term ``renormalized'' is motivated by the normalizing effect of the denominator in \autoref{eq:renormalized-barycentric-coordinates}.
While a simple subset of compositional data does not generally sum to one, $\sum_{k\in K}\pi_k \leq 1$, it is easy to show that the renormalized barycentric subset sums to one, $\sum_{k \in K}\pit_k = 1$.
Moreover, the ratio of every two renormalized coordinates $\tilde{\pi}_n, \tilde{\pi}_m$ equals the ratio of the original coordinates $\pi_n, \pi_m$ since these ratios are clearly invariant to the division by the same normalizing constant,
\begin{equation}
    \frac{\pi_n}{\pi_m}=\frac{\pi_n / \sum_{k\in K}\pi_{k}}{\pi_m / \sum_{k\in K}\pi_{k}}\defeq\frac{\tilde{\pi}_n}{\tilde{\pi}_m},
\end{equation}
as depicted in \autoref{fig:invariance_ratio}.

\paragraph{Simplicial complex}
Following the definition of \textcite{lee_introduction_2000}, a simplicial complex $\K$ is a set of simplices in an Euclidean space that satisfies (see \autoref{fig:simplicial_complex} for an illustration):
\begin{enumerate}
    \item For every simplex~$\sigma\in\K$, every face of $\sigma$ is in $\K$;
    \item the intersection of any two simplices~$\sigma_1,\sigma_2\in\K$ is either empty or a face of both $\sigma_1$ and $\sigma_2$.
\end{enumerate}
A simplicial complex~$\K$ is a \emph{pure simplicial complex} if all simplices~$\sigma\in\K$ have equal dimension.
In what follows, we will project a point~$x$ from the $(J-1)$-dimensional simplex $\Delta$ onto $J$ points $x_1,\ldots,x_J$, where each projected point $x_j$ ($j=1,\ldots,J$) lies in a $(J-2)$-dimensional simplex $\sigma_{-j}$, and all simplices $\sigma_{-1}, \ldots, \sigma_{-J}$ form a pure simplicial complex.
Since the number of lower-dimensional simplices $\sigma$ and the dimensionality of the higher-dimensional point $x$ are equal in our application, we refer to both as $J$.
Further, we will need to refer to the set $\{x_1, \ldots, x_J\}$, where each point lies in exactly one simplex $\sigma_{-j}$ of the simplicial complex~$\K=\{\sigma_{-1},\ldots,\sigma_{-J}\}$, and the indices $j$ coincide (see previous section for details on the $-j$ index notation).
We capture this with the following product notation:
\begin{equation}
    \prod\limits_{j=1}^J \sigma_{-j} = \sigma_{-1} \times \ldots \times \sigma_{-J} = \Big\{ x_j \,\Big\vert\, x_j\in\sigma_{-j} \Big\}_{j=1}^J
\end{equation}

\paragraph{Perspective Projection}
Let $\Delta$ be a ($J-1$)-simplex and $(\pi_1, \ldots, \pi_J)$ be barycentric coordinates of a point~$x\in\Delta$.
Then, we define
\begin{equation}\label{eq:perspective-projection}
    \begin{aligned}
        \psi_j: \Delta&\rightarrow \sigma_{-j}\\
        x&\mapsto \psi_j(x)\\
        (\pi_1, \ldots, \pi_J) &\mapsto 
        \left( \pit_1,\ldots, \pit_{j-1}, \pit_{j+1}, \ldots, \pit_J \right)
    \end{aligned}
\end{equation}
as the \emph{perspective projection} of $x$ about the vertex~$v_j$ onto the opposing facet~$\sigma_{-j}$.
This corresponds to shooting a ray from the vertex~$v_j$ through the point~$x$, and the intersection of that ray with the opposing edge is the image $\psi_j(x)$.
\autoref{fig:invariance_ratio} provides an illustration for a triangle (2-simplex) $\Delta$ with edges $v_1, v_2, v_3$, where the perspective projection $\psi_1$ projects the point~$x$ about the vertex~$v_1$ onto the opposing edge $\sigma_{-1}=\overline{v_2v_3}$ ($\overline{AB}$ denotes a line segment from $A$ to $B$).
It is evident that perspective projection about a vertex $v_j$ is the geometrical equivalent to renormalization (\autoref{eq:renormalized-barycentric-coordinates}) after removing the $j^{\text{th}}$ component.
For the theorems below, it is crucial that perspective projection does not affect the ratios of the remaining components' barycentric coordinates.

\section{Related Work}

\begin{figure}[t]
    \centering
        \begin{subfigure}[t]{0.38\linewidth}
            \includegraphics[width=\linewidth]{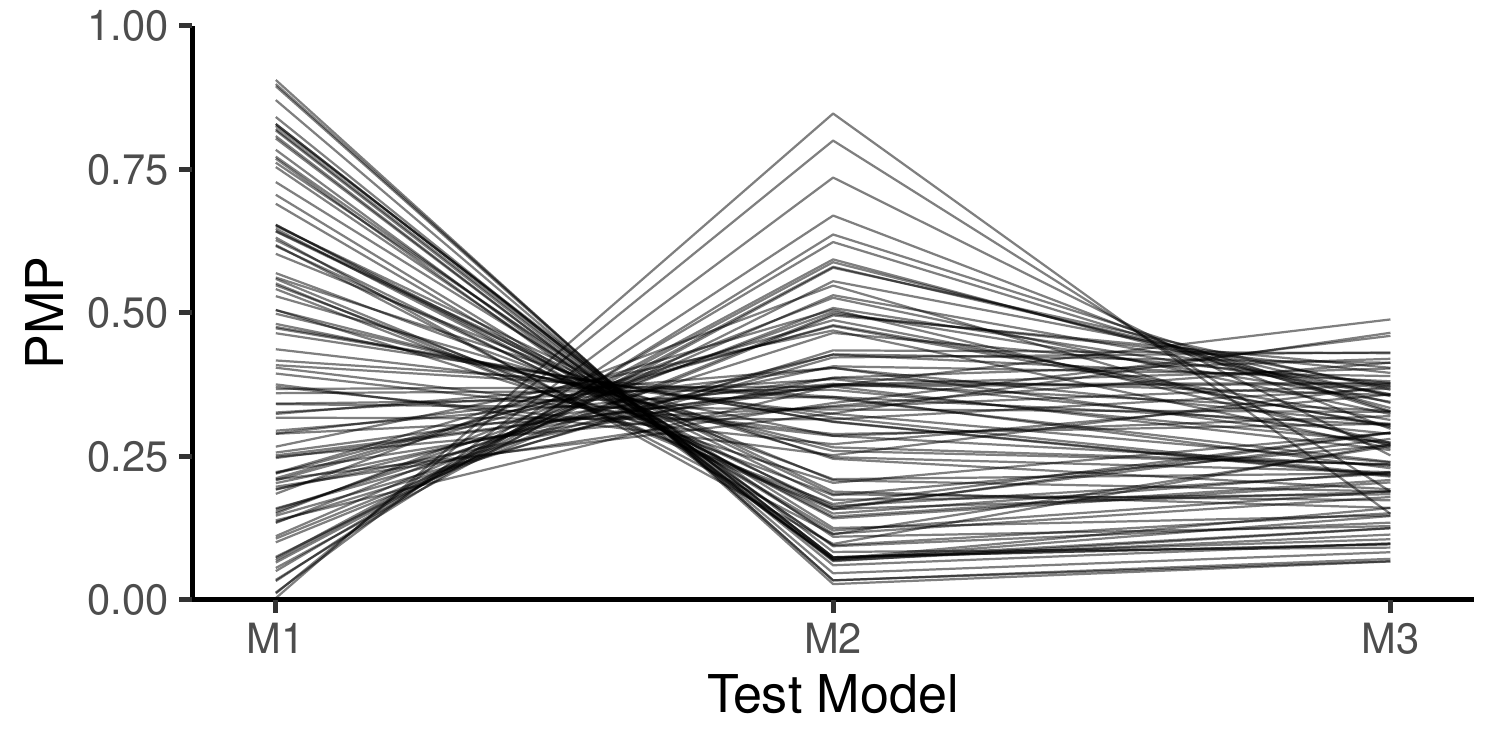}
            \caption{Parallel coordinates.}
             \label{fig:pmp-visualization-strategies:parallel-coordinates}
        \end{subfigure}
        \begin{subfigure}[t]{0.38\linewidth}
            \includegraphics[width=\linewidth]{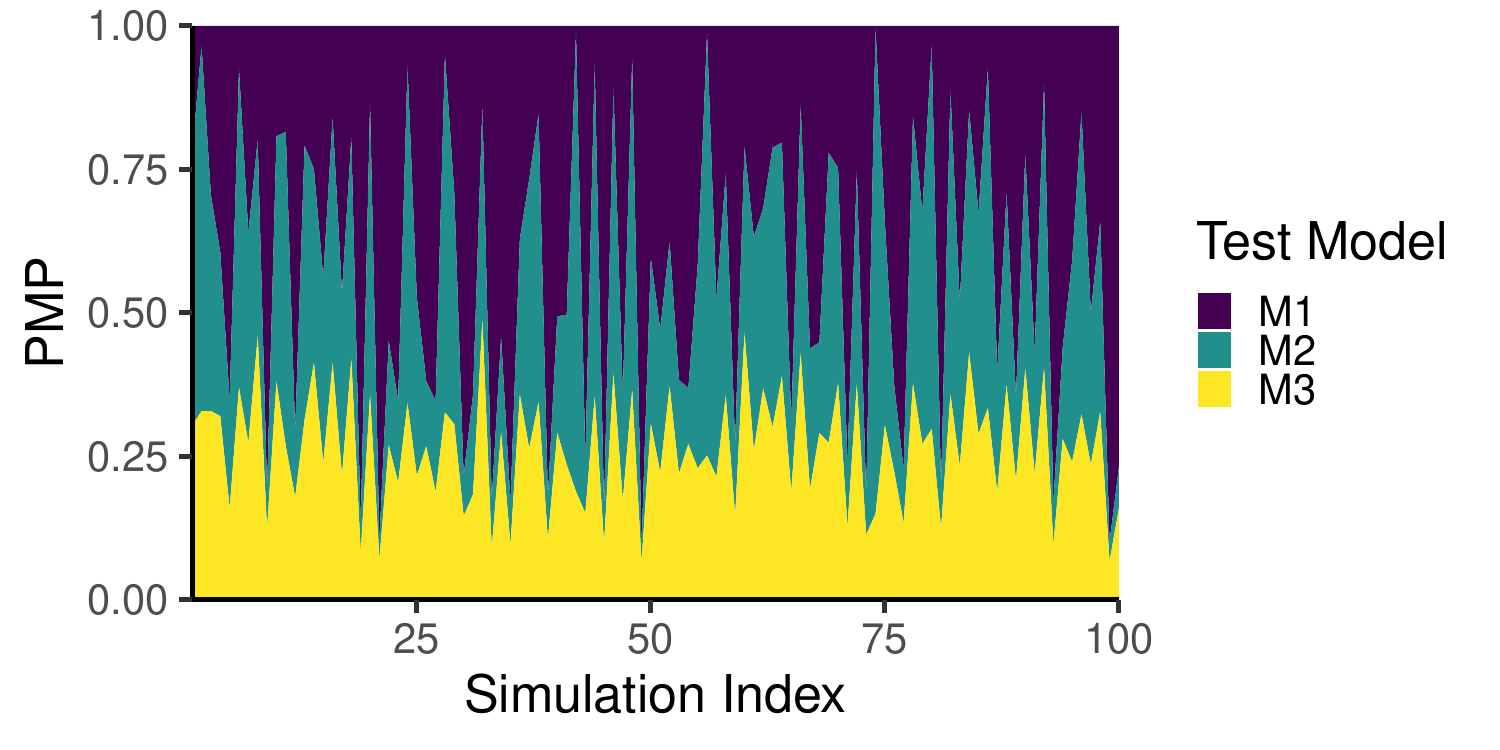}
            \caption{Stacked plot.}
            \label{fig:pmp-visualization-strategies:stacked-plot}
        \end{subfigure}
    \begin{subfigure}[t]{0.23\linewidth}
        \includegraphics[width=\linewidth]{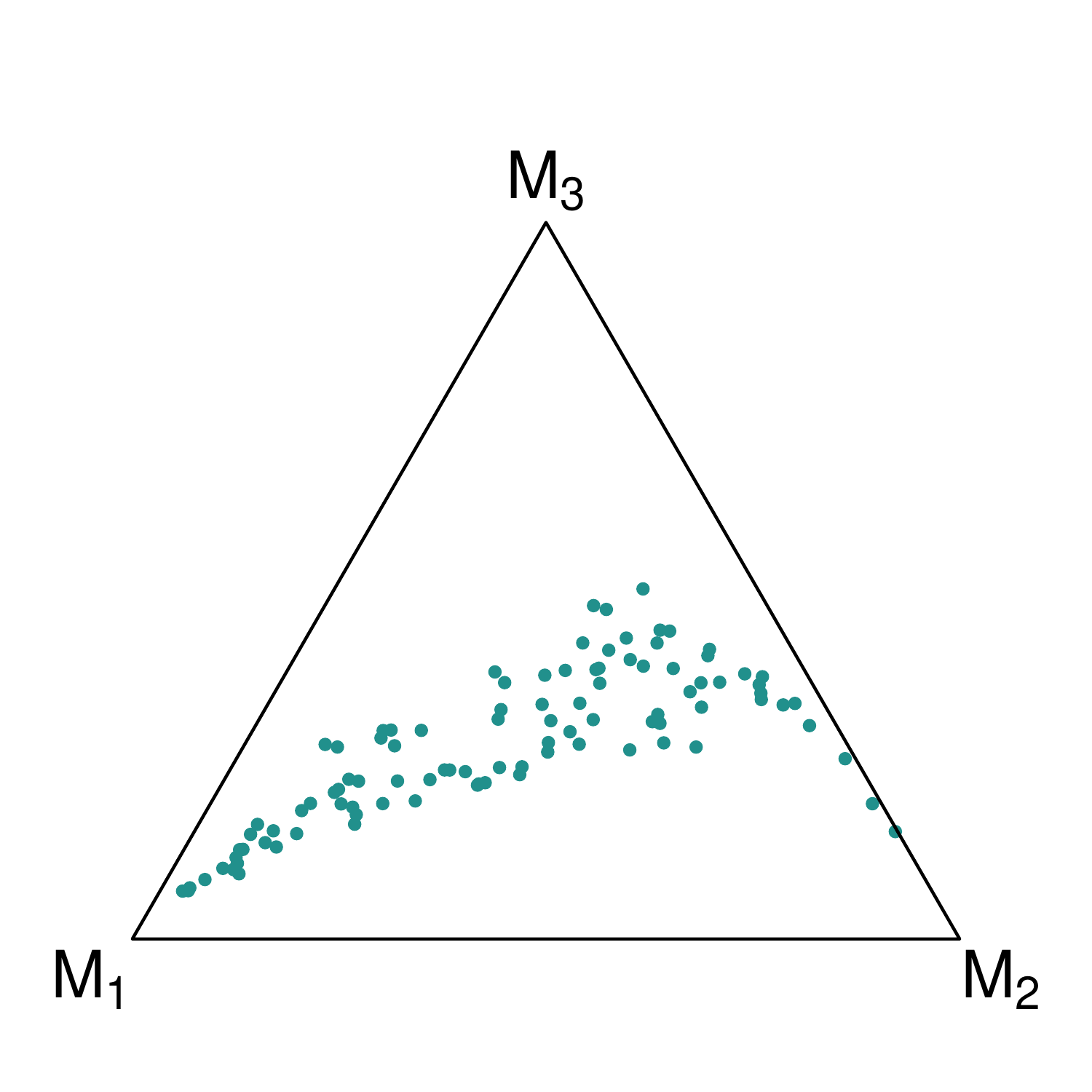}
        \caption{Simplex plot.}
        \label{fig:pmp-visualization-strategies:simplex-plot}
    \end{subfigure}
    \caption{Existing visualization techniques for compositional data. The data are equal for all three visualization methods and they stem from a Bayesian model comparison experiment: For $i=1,\ldots,100$, we have posterior model probabilities $p(M_1\given y^{(i)}), p(M_2\given y^{(i)}), p(M_3\given y^{(i)})$ which are, per construction, non-negative and sum to one for each $i$. (a) Parallel coordinates convey the dispersion for each compartment $M_j$. Here, $M_3$ has less variance across simulation indices $i$. (b) Stacked plots illustrate the dependency on a third variable. Here, there is no systematic dependency on the simulation index $i$. (c) Simplex plots visualize the relation between compartments. Here, we observe a banana-shaped relation where some combinations of compartments are more likely than others.}
    \label{fig:pmp-visualization-strategies}
\end{figure}

In the following, we will briefly compare three fundamental visualization techniques to plot compositional data: parallel coordinate plots, stacked plots, and simplex (aka ternary) plots. An illustration of each method for a fixed data set is displayed in \autoref{fig:pmp-visualization-strategies}.

\textit{Parallel coordinate plots} \autocite{klinger_parallel_1991, inselberg_parallel_2009} depict the dimension as a variable on the $x$-axis (see \autoref{fig:pmp-visualization-strategies:parallel-coordinates}).
This means that, in principle, there is no upper bound on the number of data dimensions. However, correlations and clusters are not immediately visible, which is a major conceptual drawback of parallel plots.

\textit{Stacked plots} \autocite{byron_stacked_2008} are vertically stacked area plots with a continuous $x$-axis (see \autoref{fig:pmp-visualization-strategies:stacked-plot}).
The dimension is encoded in the fill color or pattern of the respective area.
Stacked plots can communicate a relatively large number of dimensions, with an upper bound given by the number of discrete levels for the area fill (e.g., color or fill pattern).
Global trends, such as high values on one dimension throughout many data sets, are easily visible.
Furthermore, an additional (continuous) variable of interest can be plotted on the $x$-axis.
As a drawback, correlations between data instances as well as clusters are not immediately visible.
Moreover, the usability of stacked plots is influenced by crucial design choices, such as ordering \autocite{basu_aesthetics_2021}, layout \autocite{he_optimal_2022}, or type \autocite[classical, inverting, and diverging;][]{indratmo_efficacy_2018}.

\textit{Simplex plots} \autocite[aka. ternary plots;][]{howarth_sources_1996} leverage the fact that compositional data~$(\pi_1, \ldots, \pi_J)$ fulfil the properties of barycentric coordinates (they sum to 1 and are non-negative) according to \autoref{eq:simplex-definition}.
Consequently, the simplex plot visualizes the data as a point in an equilateral triangle ($2$-simplex) by interpreting the data as barycentric coordinates (see \autoref{fig:pmp-visualization-strategies:simplex-plot}).
The main advantage of simplex plots is that correlations are immediately visible.
The crucial drawback of a simplex plot is the data dimension in the visualization:
In an $n$-dimensional visualization, the conventional simplex plot is limited to $n+1$ components.
In a printed $2$D medium, this implies an upper bound of $3$ dimensions for compositional data.

Within the scope of this paper, we will extend simplex plots in order to preserve their advantages (i.e., conveying correlations between components) while pushing the envelope on their main drawback, namely the limited number of dimensions.

\begin{figure*}[t]
    \centering
    \begin{subfigure}[t]{0.78\linewidth}
       \includegraphics[width=0.9\linewidth]{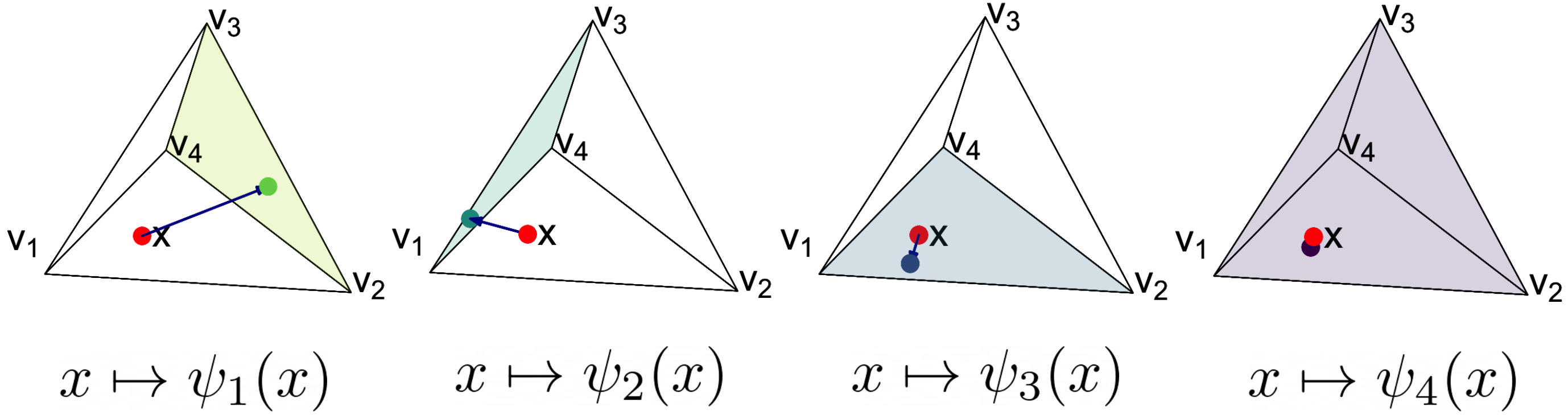}
       \caption{Projection of a single point $x$ onto each facet.}
    \end{subfigure}
    \hfill
    \begin{subfigure}[t]{0.2\linewidth}
\includegraphics[width=\linewidth]{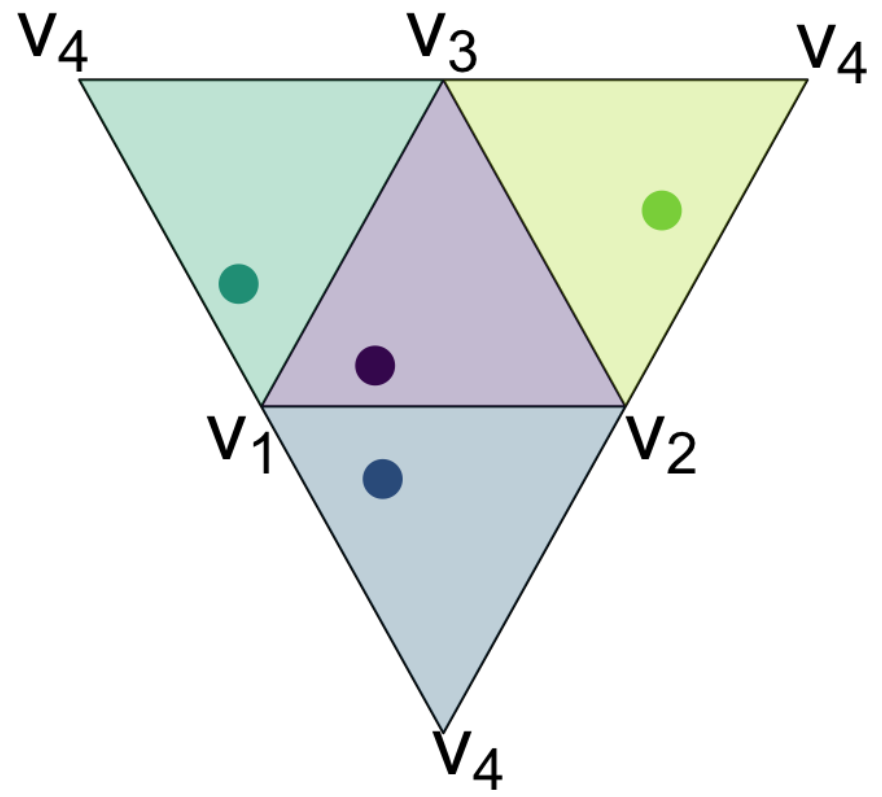}
       \caption{Unfolded projections.}
    \end{subfigure}
    \caption{Illustration of the simplex projection~$\phi=(\psi_1, \psi_2, \psi_3, \psi_4)$ for a single point~$x$ in the tetrahedron~$\Delta^3$.}
    \label{fig:preimage-matching-points-illustration}
\end{figure*}

\section{Simplex Projection}

We leverage the structure of compositional data to propose a visualization method with less image dimensions than data dimensions.
Precisely, we show that all the points in a higher dimensional simplex can be projected onto its facet without loss of information by proving that our simplex projection is bijective. Consequently, compositional data with any dimension can be projected onto a 2D canvas.
However, in this paper, we specifically highlight the case of $J = 4$ components, that is, data that could be na\"ively visualized via a tetrahedron.
After proving that the method acts as a bijective mapping between the full-order simplex and lower-order multivariate marginals for single points, we show how the method generalizes to entire sets of points and even to continuous probability density functions.
	
\subsection{Single Point}\label{sec:methods:simplex-projection:labeled-points}
Consider a point~$x\in\Delta$ in the $(J-1)$ simplex.
In the following, we will prove that a \emph{specific} set of perspective projections $\big(\psi_1,\ldots,\psi_J\big)$ constitutes an invertible mapping function $\phi$.
That is, we can reconstruct the original point $x$ from its image $\phi(x)$ under the mapping function $\phi$.
The exact form of $\phi$ is illustrated in \autoref{fig:preimage-matching-points-illustration} and formalized below.

\begin{theorem}[Bijective simplex projection for labeled points]\label{theo:preimage-matching-points}
Let $\Delta$ be a $(J-1)$-simplex, $\K=\left\{\sigma_{-j}\right\}_{j=1}^J$ be a pure simplicial complex of the facets of $\Delta$, and $\psi_j(x)$ the perspective projection of $x$ onto $\sigma_{-j}$ about the vertex~$v_j$. Further, let $\mathrm{Img}_{\phi}$ be the image of $\phi$, as detailed in Appendix \ref{app:image-domain}.
Then,
\begin{equation}\label{eq:bijection-Delta-K}
    \begin{aligned}
    \phi: \Delta &\rightarrow \mathrm{Img}_{\phi}(\Delta)\\
                x & \mapsto 
        \left(
            \psi_j\left(x\right)\in\sigma_{-j}
        \right)_{j=1}^J
\end{aligned}
\end{equation}
is a bijective mapping from the $(J-1)$-simplex~$\Delta$ to the set of compatible projections in the product set of the $(J-2)$-facets of $\Delta$.
What is more, only two matching projections $T\in\sigma_{-1}, R\in\sigma_{-2}$ onto different simplices~$\sigma_{-1}, \sigma_{-2}\in\K$ suffice to uniquely define the original point~$x\in\Delta$.
\end{theorem}
The proof of \autoref{theo:preimage-matching-points} can be found in Appendix \ref{app:proof-theorem-1}.

\subsection{Set of Points}
We previously treated the simplex projection~$\phi$ of a single point~$x\in\Delta$ and proved that $\phi$ is a bijection.
The next step generalizes the simplex projection to a \emph{set} of $L$ points~$\{x^{(1)}, \ldots, x^{(L)}\}\equiv\{x^{(l)}\}_{l=1}^L$ with shorthand notation $\{x^{(l)}\}$ when the context is clear.
One might be tempted to assume that we can simply apply $\phi$ to each point in the set individually and preserve the bijection, and this is technically true---under one critical assumption, which does not hold in practical applications: We do not know which projections match to the same original point~$x$ in the preimage, as explained more technically in the following.

For each single point~$x^{(l)}$, we can calculate $\phi(x^{(l)})$ according to \autoref{eq:bijection-Delta-K}, and we can readily apply \autoref{theo:preimage-matching-points} to recover $x^{(l)} = \phi^{-1}\big(\phi(x^{(l)})\big)$.
\autoref{fig:preimage-matching-points-illustration} illustrates this for the tetrahedron ($J=4$).
Now consider two points~$x^{(n)}, x^{(m)}$ from the preimage, and their projections~$\phi(x^{(n)}), \phi(x^{(m)})$.
In order to recover the original points with \autoref{theo:preimage-matching-points}, we need the barycentric coordinate ratios of each image, which is where the critical assumption becomes salient.
On each facet~$\sigma_{-j}$, we now have \emph{two} projections, namely $\psi_j(x^{(n)})$ and $\psi_j(x^{(m)})$.
However, the two projections are indistinguishable\footnote{Theoretically, it is possible to color all projections arising from $x^{(n)}$ in one color, all projections arising from $x^{(m)}$ in another color, and so forth, but this quickly becomes infeasible for a larger set of projected points.}, and it is not clear which one arose from $x^{(n)}$ and which one arose from $x^{(m)}$.
Yet, the argumentation in \autoref{theo:preimage-matching-points} builds on using coordinate ratios of an original point from its different projections.
Therefore, the previous proof cannot be readily applied to a set of images when the \emph{labels} (i.e., which projections arose from which original point) are unknown.

\begin{figure*}[t]
    \centering
    \begin{subfigure}[t]{0.78\linewidth}
       \includegraphics[width=0.9\linewidth]{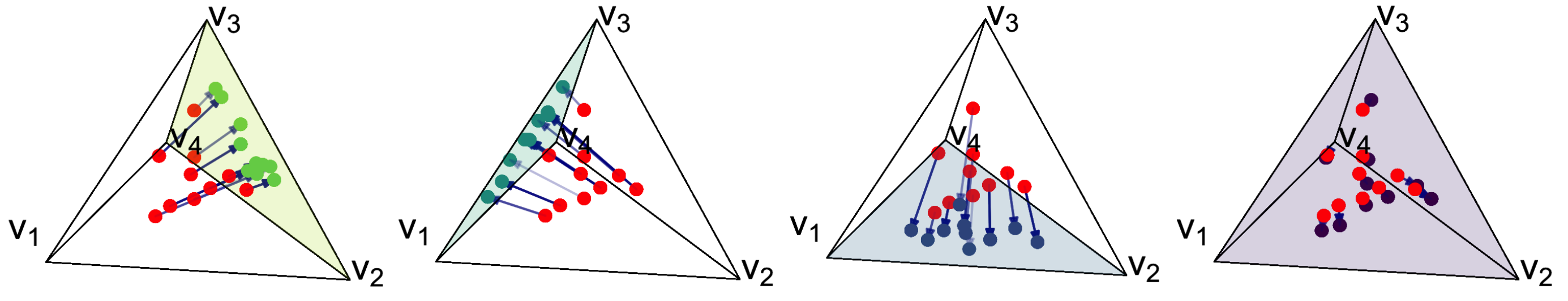}
       \caption{Projection of a set of points $\{x^{(l)}\}$ onto each facet.}
    \end{subfigure}
    \hfill
    \begin{subfigure}[t]{0.2\linewidth}
\includegraphics[width=\linewidth]{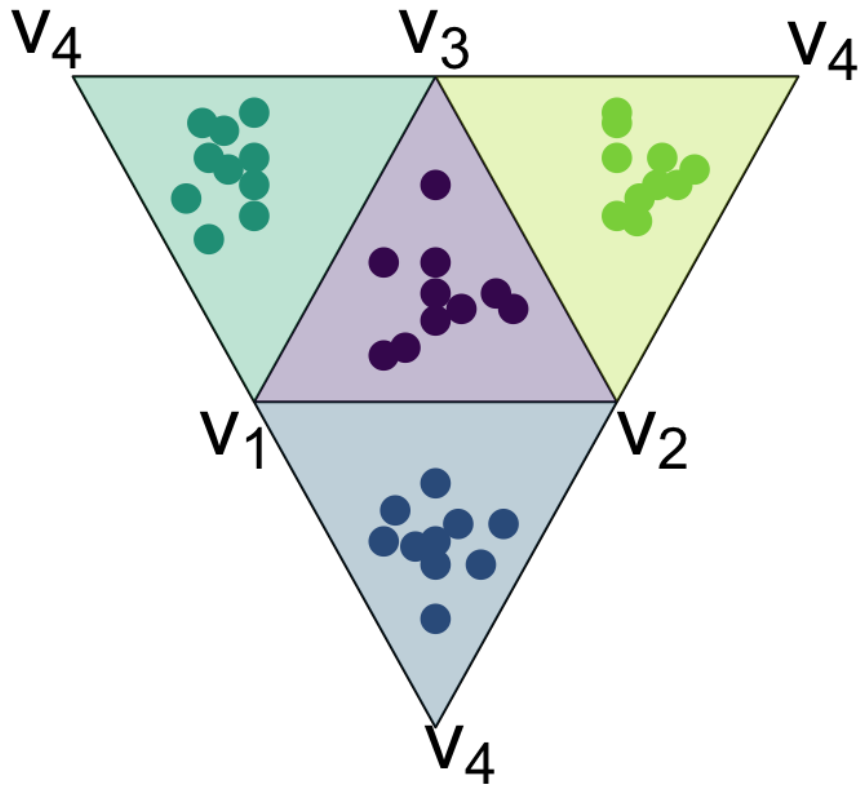}
       \caption{Unfolded projections.}
    \end{subfigure}
    \caption{Illustration of the simplex projection $\Phi$ for a set of points $\{x^{(l)}\}$.
The labels of the projections onto the facets in \textbf{(c)} are lost: We do not know which points across facets originate from the same original point in the preimage. Yet, \autoref{theo:preimage-unlabeled} shows that the original points in \textbf{(a)} can still be recovered from the unlabeled projections in \textbf{(c)}}.
    \label{fig:preimage-set-points-illustration}
\end{figure*} 

As an extension to \autoref{theo:preimage-matching-points}, we define a function~$\Phi$ which applies the simplex projection~$\phi$ to each element of a \emph{set} of $L$ points $\{x^{(l)}\}$,
\begin{equation}
    \Phi
        \left(
            \left\{
                x^{(l)}
            \right\}%
        \right) =
        \left\{
                \phi(x^{(l)})
        \right\}%
,
\end{equation}
  where we use the shorthand set notation $\{\cdot\}$ for brevity.
If the point labels $l=1,\ldots, L$ are known for all images (up to permutation), the function~$\Phi$ will be invertible as well because $\phi^{-1}$ is invertible:
\begin{equation}\label{eq:proof-sets-equality-theo1}
    \begin{aligned}
        \Phi^{-1}
        \left(
            \left\{
                \phi(x^{(l)})
            \right\}%
        \right)
        &=  
        \left\{
            \phi^{-1}\left(
                \phi(x^{(l)})
            \right)
        \right\}%
        =
        \left\{
            x^{(l)}
        \right\}%
        .
    \end{aligned}
\end{equation}

Consequently, it remains to be proven that the labels across projections onto different facets are unique (up to permutations of labels) such that the problem can be reduced to \autoref{theo:preimage-matching-points}.
Since performing a perspective projection on a point~$\Delta^0 (J=1)$ is not sensible and a perspective projection of points on a line~$\Delta^1 (J=2)$ would project all points to a single renormalized coordinate~$\pit_j=1$, we formulate the following theorem for $J\geq 3$.
		
\begin{theorem}[Bijective simplex projection for sets of points]\label{theo:preimage-unlabeled}
Let $\Delta$ be a $(J-1)$-simplex ($J\geq 3$), $\K=\left\{\sigma_{-j}\right\}_{j=1}^J$ be a pure simplicial complex of the facets of $\Delta$, $\phi$ a bijective mapping (\autoref{eq:bijection-Delta-K}) with inverse function~$\phi^{-1}$ and $\psi_j(x)$ the perspective projection of $x$ onto $\sigma_{-j}$ about the vertex~$v_j$. Further, let $\mathrm{Img}_{\phi}$ denote the image of $\phi$, as described in Appendix \ref{app:image-domain}.
Then,
\begin{equation}\label{eq:bijection-Delta-K-set}
    \begin{aligned}
    \Phi: \left\{\Delta^{(l)}\right\}_{l=1}^L &\rightarrow \left\{
    \mathrm{Img}_{\phi}(\Delta^{(l)})
    \right\}_{l=1}^L\\
    \left\{x^{(l)}\right\}_{l=1}^L & \mapsto 
        \left\{
            \phi\left(
                x^{(l)}
            \right)
        \right\}_{l=1}^L %
\end{aligned}
\end{equation}
is a bijective mapping.%
\end{theorem}
The proof of \autoref{theo:preimage-unlabeled} is presented in Appendix~\ref{app:proof-theorem-2}.

\begin{figure}[t]
\begin{minipage}{0.45\linewidth}
    \begin{algorithm}[H]
    \caption{Marginal density approximation}
    \label{alg:simplex-marginal-density-numeric}
    \begin{algorithmic}[1]
        \REQUIRE{Simplex $\Delta$, probability density function $p(x)$ over $\Delta$, integral accuracy $M$}
        \ENSURE{Approximate marginal density $\hat{p}(z)$}
        \FOR{$\sigma_{-j}\in\K$}
            \FOR{$z\in\subdivision(\sigma_{-j})$}
                \STATE{$s = \overline{z v_j}$}
                \STATE{$\hat{p}(z)=0$}
                \FOR{$m=1,\ldots, M$}
                    \STATE{$x_m\leftarrow z + \frac{s}{M}$}\hfill\COMMENT{equidistant points on $s$}
                    \STATE{$\hat{p}(z) \mathrel{+}= p(x_m)\,\frac{|s|}{M}$}
                \ENDFOR
            \ENDFOR
        \ENDFOR
        \RETURN $\hat{p}(z)$
    \end{algorithmic}
    \end{algorithm}
\end{minipage}
\hfill
\begin{minipage}{0.53\linewidth}
    \begin{figure}[H]
        \centering
        \includegraphics[width=0.9\linewidth]{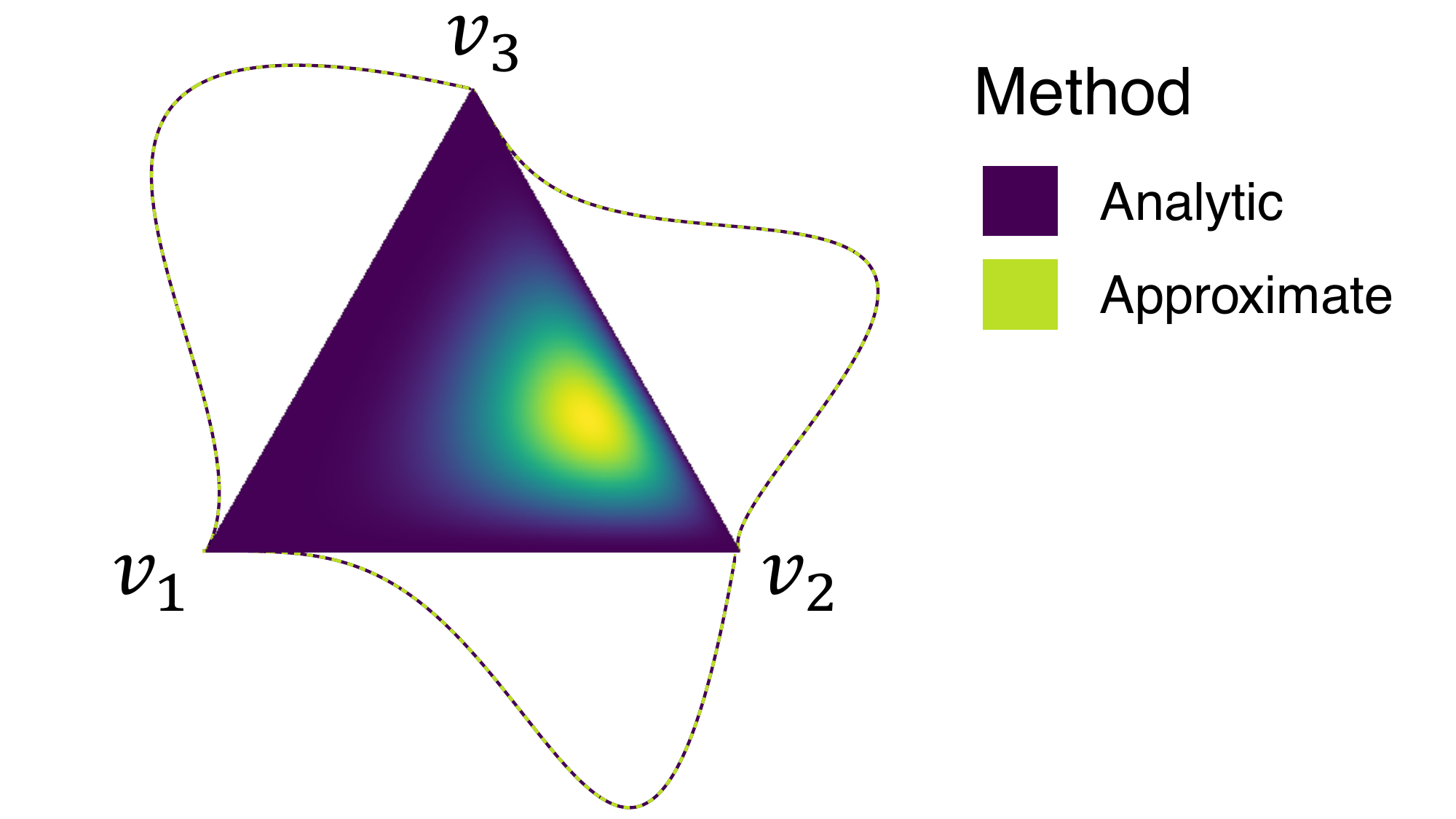}
        \vspace*{-0.30cm}
        \caption{Validation of the marginal density approximation: For the joint Dirichlet density over $\Delta^2$ (triangle interior), the $2$-variate marginal distributions are approximated (dashed green) and plotted against the analytic ground-truth density (solid purple) orthogonal to the respective edge.
        Our approximation is indistinguishable from the analytic ground-truth density.}
        \label{fig:validation-numeric-approximation}
    \end{figure}    
\end{minipage}
\end{figure}

\subsection{Continuous Probability Densities}
In what follows, we will extend our method from discrete sets of points to continuous densities, such as probability density functions over the simplex~$\Delta$.
Given a density $p(x)$ defined over the simplex~$\Delta$, we can approximate the (multivariate) marginal densities of the projections in the simplices of the simplicial complex~$\K$ numerically (see Algorithm~\ref{alg:simplex-marginal-density-numeric}).
This approach is conceptually similar to pre-integration \autocite{griewank_high_2018} for each component, but offers a strong statistical foundation through the principled statistically equivalent operation of \emph{marginalization}.
Furthermore, the simplex projection method is information-preserving in finite dimensions (as opposed to pre-integration in general), as shown in \autoref{theo:preimage-matching-points} and \autoref{theo:preimage-unlabeled}.

Let $\Delta$ be a $(J-1)$-simplex, $p(x)$ be a probability density function over $\Delta$, and $\K=\left\{\sigma_{-j}\right\}_{j=1}^J$ be a pure simplicial complex of the facets of $\Delta$.
For each simplex~$\sigma_{-j}\in\K$, we define a node-based subdivision~$\subdivision(\sigma_{-j})$ of depth $D$.

The depth $D$ controls the number of nodes and thus the resolution of the approximated marginal density.
For each node~$z$ of the subdivision~$\subdivision(\sigma_{-j})$, we construct a line segment to the opposing vertex~$s=\overline{z v_j}$ in the space of the original higher-order simplex~$\Delta$.
Then, we choose $M$ equidistant points~$(x_m)_{m=1}^M, x_m\in\mathbb{R}^{J-1}$ on the line segment~$s$ and evaluate the density~$p(x_m)$ for each point~$x_m$.
The line integral of $p(x)$ along $s$ is then approximated by
\begin{equation}\label{eq:line-integral-approximation}
    \int_{s} p(x)\diff x = \lim_{M\rightarrow \infty}
    \sum\limits_{m=1}^{M} p(x_m)\,\frac{|s|}{M}
\end{equation}
with integration step size~$\delta = |s|/M$, where $|s|$ is the length of the line segment $s$.
In the implementation, $M$ is finite and acts as a hyperparameter which controls the accuracy of the approximation in \autoref{eq:line-integral-approximation} through the number of equidistant points~$x_m$.

In the software implementation, the density at the bounds $m=1$ and $m=M$ in \autoref{eq:line-integral-approximation} might be undefined.
We tackle this by setting $p(x_1)=p(x_M)=0$, and the de-facto effect of this practical adjustment vanishes for increasing accuracy~$M$.

\paragraph{Proof-of-Concept with an Analytic Density}
We validate our approximation for a Dirichlet distribution which has known analytic marginal distributions as a ground-truth. 
For $(\pi_1, \ldots, \pi_J)\in\Delta^{J-1}, J\geq2$, the Dirichlet distribution $\mathrm{Dir}(\alpha_1,\ldots,\alpha_J)$ has the probability density function
\begin{equation}
\begin{aligned}
    p(\pi_1,\ldots,\pi_J\given\alpha_1,\ldots,\alpha_J) &= \mathbf{B}(\alpha_1,\ldots,\alpha_J)^{-1}\,\prod\limits_{j=1}^J\pi_j^{\alpha_j-1}, \alpha_j>0\,\forall\,j=1,\ldots,J\;\;\text{with}\\
    \mathbf{B}(\alpha_1,\ldots,\alpha_J)&=\left(\prod\limits_{j=1}^{J}\Gamma(\alpha_j)\right)\,\Gamma\left(\prod\limits_{j=1}^J \alpha_j\right)^{-1},
\end{aligned}
\end{equation}
where $\Gamma$ denotes the Gamma function. The Dirichlet distribution is a multivariate generalization of the Beta distribution \autocite{kotz_continuous_2000}.
To ease the illustration, we study a Dirichlet distribution $\mathrm{Dir}(\alpha_1, \alpha_2,\alpha_3)$ over the 2-simplex~$\Delta^2$ with parameter vector $(\alpha_1, \alpha_2, \alpha_3)=(2,5,3)$.
This implies the following analytic multivariate marginal distributions \autocite{aitchison_statistical_1982}, which we will use as a ground-truth to benchmark our approximation against:
\begin{equation}
    \begin{aligned}
        (\pi_1, \pi_2) &\sim \Dir(\alpha_1, \alpha_2)=\Dir(2, 5),\\
        (\pi_1, \pi_3) &\sim \Dir(\alpha_1, \alpha_3)=\Dir(2, 3),\\
        (\pi_2, \pi_3) &\sim \Dir(\alpha_2, \alpha_3)=\Dir(5, 3).
    \end{aligned}
\end{equation}
\autoref{fig:validation-numeric-approximation} illustrates that the proposed approximation technique yields essentially equal results to the analytic marginal distributions for a subgrid depth of $D=10$ ($2^{10}=1024$ evaluation nodes on each edge) and an integration accuracy of $M=1000$.
For the practical purpose of visualizing (probability) densities over higher-order simplices in a lower-order canvas, this shows that (i) the density information can be preserved through the simplex projection; and (ii) the necessary numerical approximation does not introduce substantial inaccuracies.

\begin{figure*}[t]
\centering
\begin{subfigure}[t]{0.28\linewidth}
    \includegraphics[width=\linewidth]{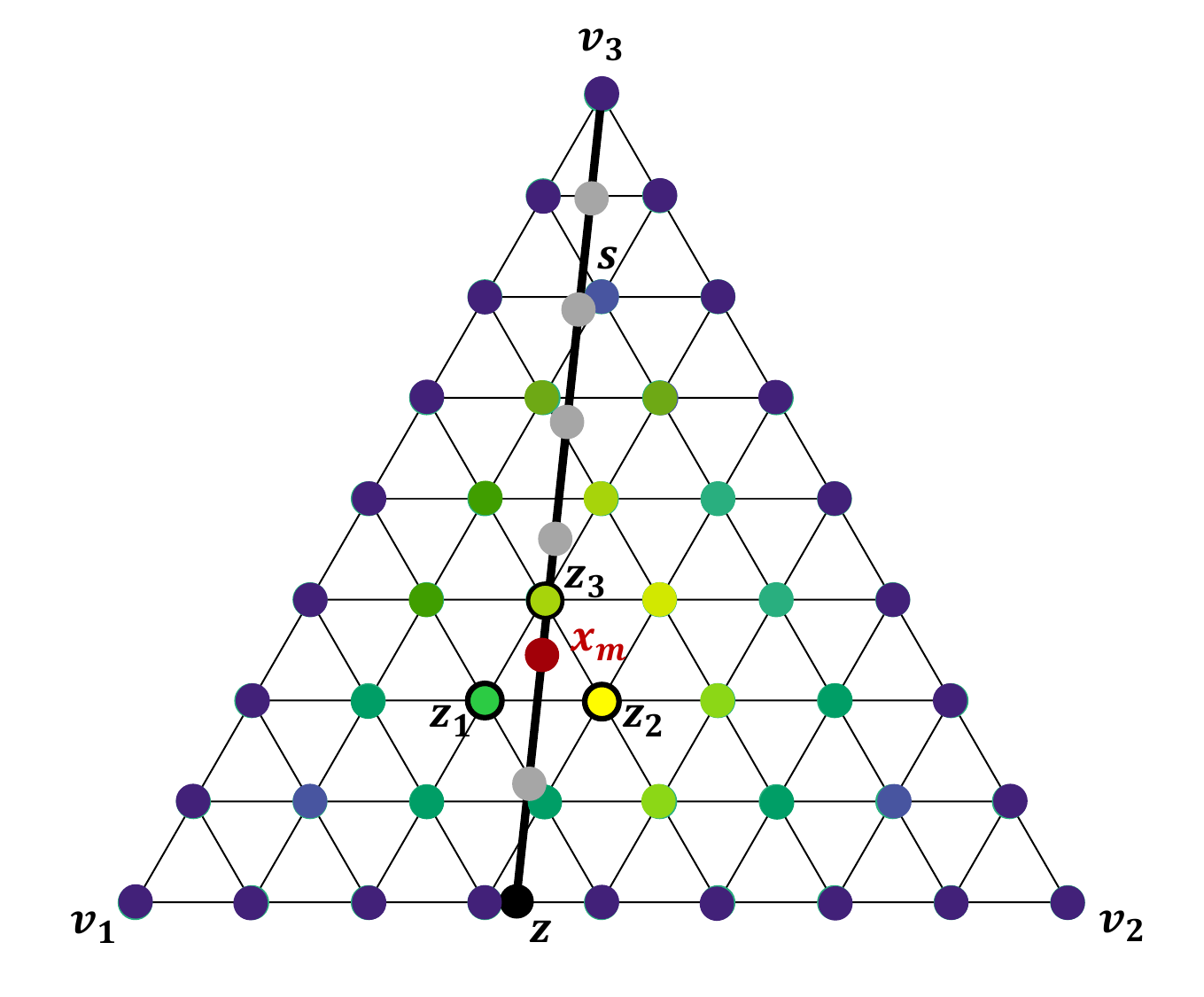}
    \caption{$x_m$ is between nodes}
\end{subfigure}
\hfill
\begin{subfigure}[t]{0.28\linewidth}
    \includegraphics[width=\linewidth]{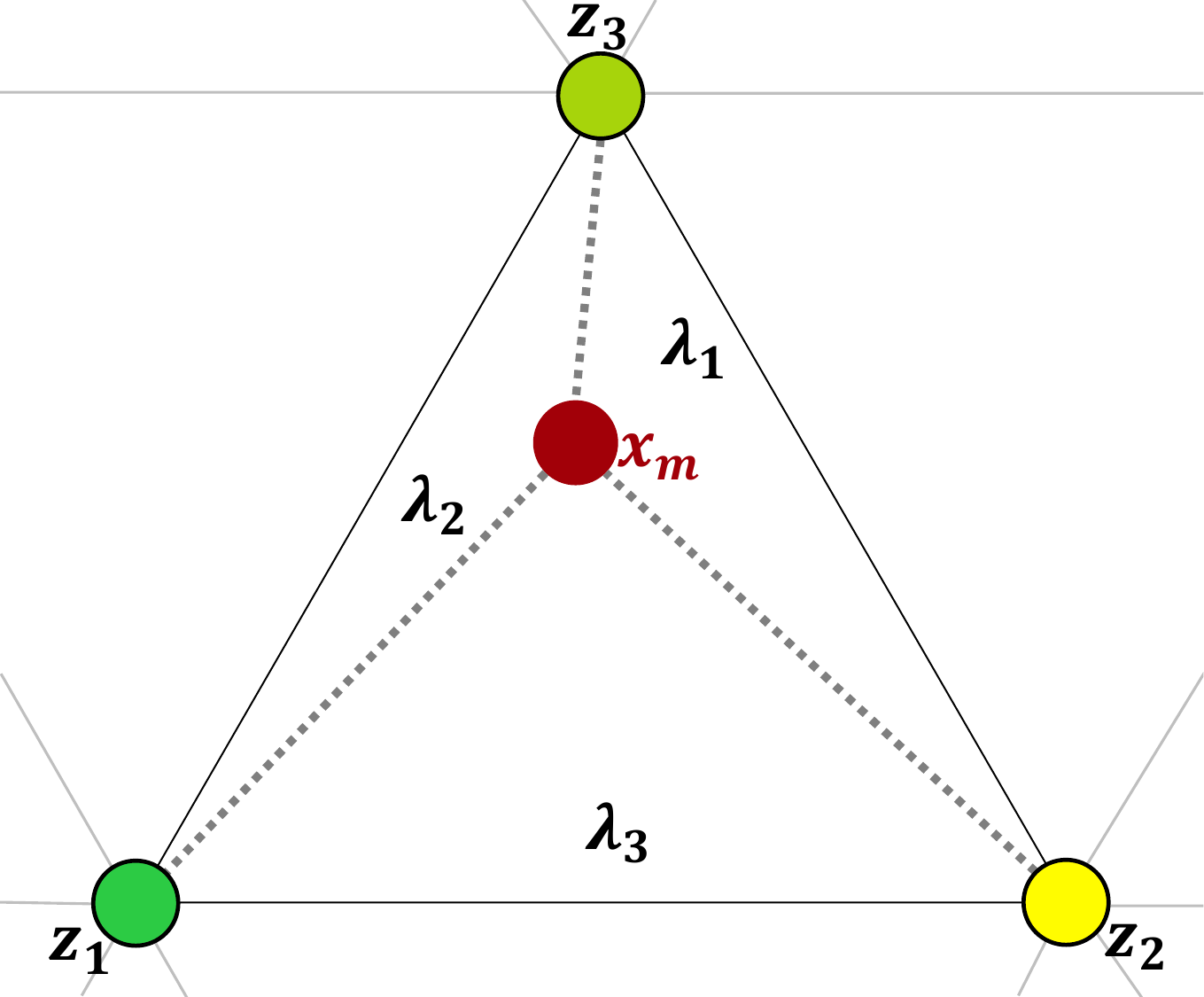}
    \caption{Barycentric approximation}
\end{subfigure}
\hfill
\begin{subfigure}[t]{0.40\linewidth}
    \centering
    \includegraphics[width=\linewidth]{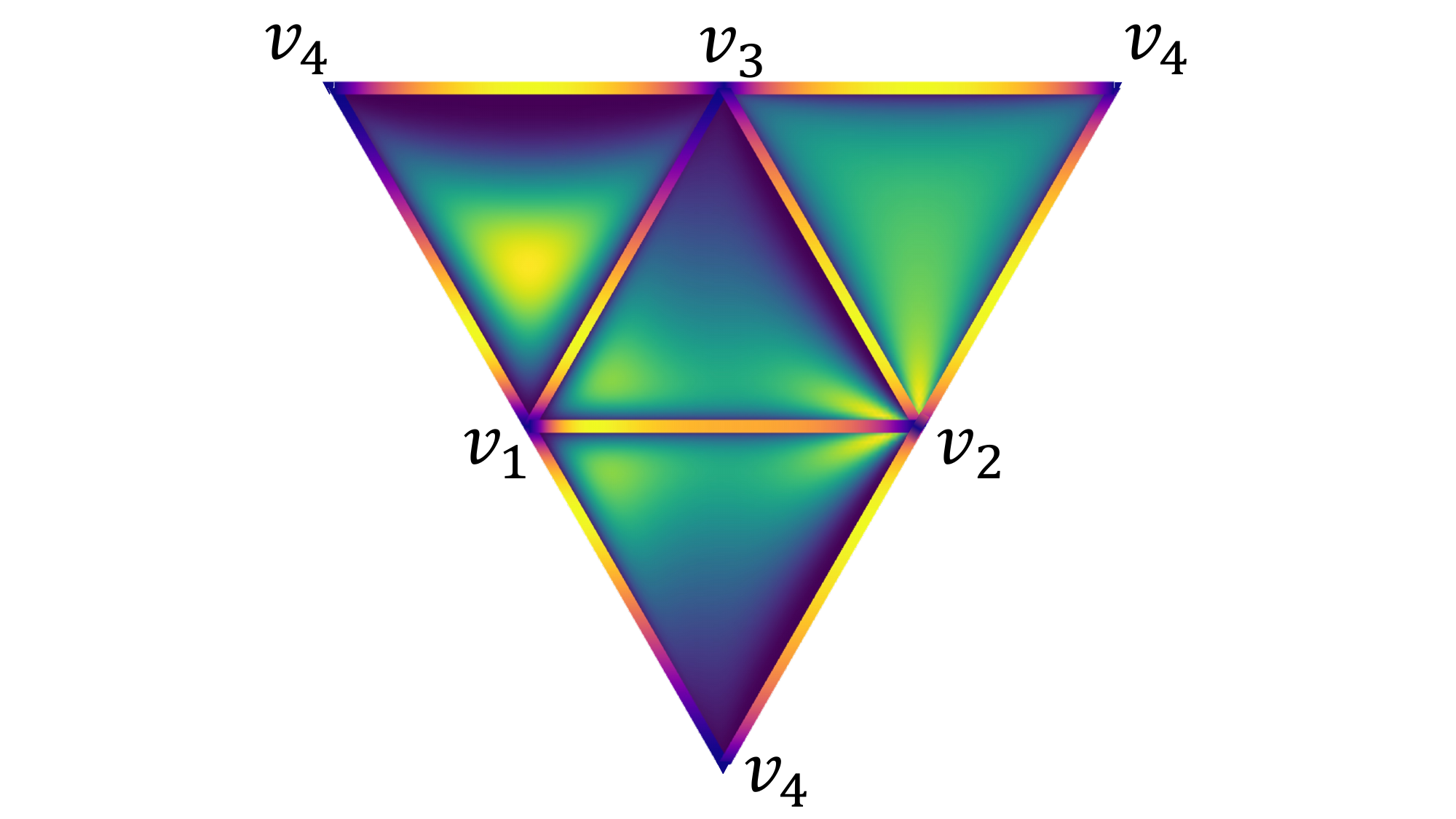}
    \caption{Recursive density approximation}
    \label{fig:numeric-approximation-recursive}
\end{subfigure}
\caption{Illustration of the recursive method to interpolate lower-order marginal densities from a joint density over a $3$-simplex (tetrahedron). \textbf{(a)} Since $p(x_m)$ cannot be evaluated directly, the density will be approximated through the $J-1=3$ nearest neighbors $z_1, z_2, z_3$. \textbf{(b)} The density of $x_m$ is approximated through barycentric interpolation of the densities of the nearest neighbor vertices $z_1, z_2, z_3$. \textbf{(c)} Applying Algorithm~\ref{alg:simplex-marginal-density-numeric} recursively yields approximations for lower-order marginal distributions as recursive interpolations. The analytic density is defined over the tetrahedron $\Delta^3$. The $3$-variate marginal density approximations $\hat{p}(\pi_1,\pi_2,\pi_3)\approx \int p(\pi_1, \pi_2, \pi_3, \pi_4)\diff\pi_4$ are depicted on the area of the triangular facets using the viridis color palette. The subsequent $2-$variate marginal density approximations are defined over the edges. These densities are obtained via recursive approximation (see \textbf{(a)} and \textbf{(b)}) and plotted on the edges with the viridis-plasma palette. For instance, the edge $\overline{v_1v_2}$ shows the recursive approximation of the marginal density $\hat{p}(\pi_1,\pi_2)\approx \int \hat{p}(\pi_1, \pi_2, \pi_3)\diff \pi_3$.}
\label{fig:numeric-approximation-recursive-concept}
\end{figure*}

\paragraph{Recursive Marginal Approximations are Possible through Interpolation}
For $J=4$ dimensions, the full-order simplex~$\Delta$ is a tetrahedron.
Accordingly, the $3$-variate marginal density distributions are defined over the $2$-simplices (triangles) at the facets of the tetrahedron, and approximated as described above.
In theory, this algorithm could be carried out again in order to obtain $2$-variate marginal distributions along the edges of each $2$-simplex.
However, Algorithm~\ref{alg:simplex-marginal-density-numeric} assumes that we can readily evaluate the density $p(x)$ at any point $x$ in the $2$-simplex (triangle).
This is not generally the case because we only have access to a numeric density for the subgrid~$Z\subseteq\Delta$ which we approximated via Algorithm~\ref{alg:simplex-marginal-density-numeric}.

We address the problem of accessing a density $p(x)$ for arbitrary $x$ through a straightforward barycentric interpolation, as explained in the following.
First, we apply Algorithm~\ref{alg:simplex-marginal-density-numeric} with the recursion step $\Delta\leftarrow\sigma_{-j}$ and proceed as usual until we need to evaluate the density $p(x_m)$ at the node $x_m$.
Because we do not generally have access to the density $p(x_m)$, we approximate $p(x_m)$ with an interpolation from its $J-1$ nearest neighbors~$z_1, \ldots, z_{J-1}$ (i.e., the vertices of the enclosing $J-2$ simplex of the subdivision, see \autoref{fig:numeric-approximation-recursive}).
The notation $z_j$ emphasizes that the density~$\hat{p}(z_j)$ of these vertices has already been approximated through the previous iteration of Algorithm~\ref{alg:simplex-marginal-density-numeric}.

With $\lambda_1, \ldots, \lambda_{J-1}$ as the barycentric coordinates of $x_m$ with reference vertices~$z_1, \ldots, z_{J-1}$, we obtain the linear barycentric interpolation $\hat{p}(x_m)~=~\sum_{j=1}^{J-1}\lambda_j\hat{p}(z_j)$ for the probability density of $x_m$. More sophisticated interpolation schemes are possible as well but do not fall within the scope of this paper.
This process can be repeated recursively (with recursion $\Delta\leftarrow\sigma_{-j}$) until $\Delta=\Delta^1$ (line segment) in Algorithm~\ref{alg:simplex-marginal-density-numeric}.
The result of applying the recursive approximation technique to an arbitrary density over the $3$-simplex is illustrated in \autoref{fig:numeric-approximation-recursive}.
	
\section{Conclusion}
Current visualization techniques for compositional data on a 2D canvas are either (i) limited to 3D data (simplex plots); or (ii) they fail to express structures across dimensions, such as correlations (parallel coordinates, stacked plots).
In this work, we overcome the limited expressiveness (i) of simplex plots while preserving the structure capturing property (ii).
Exploiting the inherent structure of compositional data, we proposed a mathematically sound perspective projection approach which corresponds to visualizing marginal densities in statistics.
The resulting visualization enjoys the remarkable property that the original joint data distribution can be reconstructed without loss of information, which we proved mathematically.
While our mathematical proof holds for an arbitrary number of dimensions, we focused our novel visualization method to representing 4D compositional data on a 2D canvas.
Future research can aim to push the envelope and propose structure- and information-preserving visualizations for higher-dimensional compositional data on a low-dimensional canvas.

\FloatBarrier
\section*{Acknowledgments}
We thank Maximilian Scholz and Javier Enrique Aguilar for stimulating discussions that played a pivotal role in transforming this method from Cthulhu triangles into a mathematically sound visualization technique.
We thank Egzon Miftari for helpful advice and feedback on mathematical intricacies of the early conceptualization.
MS thanks the Cyber Valley Research Fund (grant number: CyVy-RF-2021-16) and the ELLIS PhD program for support.
MS and PCB were supported by the Deutsche Forschungsgemeinschaft (DFG, German Research Foundation) under Germany’s Excellence Strategy -- EXC-2075 - 390740016 (the Stuttgart Cluster of Excellence SimTech).

\printbibliography

\clearpage
\appendix
\onecolumn
\begin{center}
    \Huge\scshape Appendix
\end{center}
\FloatBarrier

\section{Further Details and Proofs}
\subsection{Image of the Perspective Projection $\phi$}\label{app:image-domain}

\paragraph{The ``is compatible with'' Relation $\rel$}
		Let $\K=\left\{\sigma_1,\ldots,\sigma_J\right\}$ be a pure simplicial complex, and $I(\sigma_j)$ be the set of vertex indices of the simplex~$\sigma_j$.
		The \emph{``is compatible with'' relation}
		    \begin{equation}
		        \begin{aligned}
		    \rel\;=
		    \Big\{ (A, B)\;\Big\vert\;&
		    A=(\alpha_1, \ldots, \alpha_n)\in\sigma_1,
		    B=(\beta_1, \ldots, \beta_m)\in\sigma_2,\\
		    &\text{for}\; U = I(\sigma_1)\cap I(\sigma_2):
		    (\tilde{\alpha}_k)_{k\in U} = (\tilde{\beta}_k)_{k\in U}
		    \Big\}
		    \subseteq\sigma_1\times\sigma_2
		        \end{aligned}
		    \end{equation}
        describes equivalence of points across simplices~$\sigma_1, \sigma_2\in\K$ with respect to the ratio of their shared components, which we call ``compatibility''.
        According to conventions, we define the infix notation $A\rel B \Longleftrightarrow (A, B)\in\;\rel$.

    	\begin{figure}[t]
    		\centering
    		\begin{adjustbox}{width=0.5\textwidth}
    			\pgfdeclarelayer{bg}    %
\pgfsetlayers{bg,main} 

\begin{tikzpicture}[
	dot/.style={draw, inner sep=0pt, minimum size=6pt, circle, fill=black},
	grid/.style={draw, lightgray, opacity=0.3}
	]

	\node[draw,
	shape border rotate=120,
	regular polygon,
	regular polygon sides=3,
	minimum size =4cm] (T123) at (0,0){};
	
	\node[draw,
	shape border rotate=60,
	regular polygon,
	regular polygon sides=3,
	anchor=corner 1,
	rotate=0,
	minimum size =4cm] (T124) at (T123.corner 1){};
	
	\node[draw,
	shape border rotate=60,
	regular polygon,
	regular polygon sides=3,
	anchor=corner 2,
	rotate=0,
	minimum size =4cm] (T234) at (T123.corner 2){};
	
	\node[draw,
	shape border rotate=180,
	regular polygon,
	regular polygon sides=3,
	anchor=corner 1,
	rotate=0,
	minimum size =4cm] (T134) at (T123.corner 1){};
	
	\coordinate (T123_M1) at (T123.corner 1);
	\coordinate (T123_M2) at (T123.corner 2);
	\coordinate (T123_M3) at (T123.corner 3);
	
	\coordinate (T134_M1) at (T134.corner 1);
	\coordinate (T134_M3) at (T134.corner 2);
	\coordinate (T134_M4) at (T134.corner 3);
	
	\coordinate (T124_M1) at (T124.corner 1);
	\coordinate (T124_M4) at (T124.corner 2);
	\coordinate (T124_M2) at (T124.corner 3);
	
	\coordinate (T234_M3) at (T234.corner 1);
	\coordinate (T234_M2) at (T234.corner 2);
	\coordinate (T234_M4) at (T234.corner 3);

	\node[circle, label=below left:$v_1$] at (T123.corner 1) {};
	\node[circle, label=below right:$v_2$] at (T123.corner 2) {};
	\node[circle, label=above:$v_3$] at (T123.corner 3) {};
	
	\node[label=below:$v_4$] at (T124_M4) {};
	\node[label=left:$v_4$]  at (T134_M4) {};
	\node[label=right:$v_4$] at (T234_M4) {};

	\begin{pgfonlayer}{bg}
		
		\foreach \k in {0.1, 0.2, 0.3, 0.4, 0.5, 0.6, 0.7, 0.8, 0.9}{

			\draw [grid] let \p1=(T123_M1), \p2=(T123_M2), \n1={veclen(\x2-\x1,\y2-\y1)} in ($(T123_M1)+\k*(T134_M4)-\k*(T134_M1)$) arc (120:300:\k*\n1);
			\draw [grid] let \p1=(T123_M1), \p2=(T123_M2), \n1={veclen(\x2-\x1,\y2-\y1)} in ($(T123_M2)+\k*(T124_M4)-\k*(T124_M2)$) arc (240:420:\k*\n1);
			\draw [grid] let \p1=(T123_M1), \p2=(T123_M2), \n1={veclen(\x2-\x1,\y2-\y1)} in ($(T123_M3)+\k*(T234_M4)-\k*(T234_M3)$) arc (0:180:\k*\n1);
			
		}

		\foreach \triangle in {T123, T124, T134, T234}{
			
			\coordinate (V1) at (\triangle.corner 1);
			\coordinate (V2) at (\triangle.corner 2);
			\coordinate (V3) at (\triangle.corner 3);
			
			\foreach \k in {0.1, 0.2, 0.3, 0.4, 0.5, 0.6, 0.7, 0.8, 0.9}{
				\pgfmathsetmacro\x{\k}
				\pgfmathsetmacro\y{1.0-\x}
				\draw[grid] (barycentric cs:V1=\x,V2=\y) -- (V3);
				\draw[grid] (barycentric cs:V1=\x,V3=\y) -- (V2);
				\draw[grid] (barycentric cs:V2=\x,V3=\y) -- (V1);
			}
		}
		
	\end{pgfonlayer}{bg}

	\node[dot, color1, label=$T$] (T) at (barycentric cs:T123_M1=0.15,T123_M2=0.3,T123_M3=0.55) {};
    \path [name path=M1T] (T123_M1) -- ($(T123_M1)!7cm!(T)$);

    \path [name path=M2T] (T123_M2) -- ($(T123_M2)!7cm!(T)$);
    \path [name path=M1M3] (T123_M1) -- (T123_M3);
    \node [dot, name intersections={of=M2T and M1M3, by=R'}, label=$R'$, black] (R') at (intersection-1) {};
    \path[draw, name path=M2R', label=$t$, color1] (T123_M2) -- (R') node[midway,below] {$t$};
    \path[draw, name path=R'M4, color1] (R') -- (T134_M4) node[midway,below] {r};

\end{tikzpicture}
    		\end{adjustbox}
    	\caption{If two points are compatible according to the relation~$\rel$, they have identical projections onto shared faces.
    	$\forall R\in r:T\rel R$ because the projections on the shared face~$\overline{v_1v_3}$ are equal, $\psi_2(T)=\psi_4(R)=R'$.
    	What is more, we can extend this to all points in the simplex~$\sigma_{-4}$ which project to $R'$ as well, $\forall\,R\in r\;\forall\,\tilde{T}\in t:R\rel\tilde{T}$.}
    	\label{fig:is-compatible-with}
    	\end{figure}
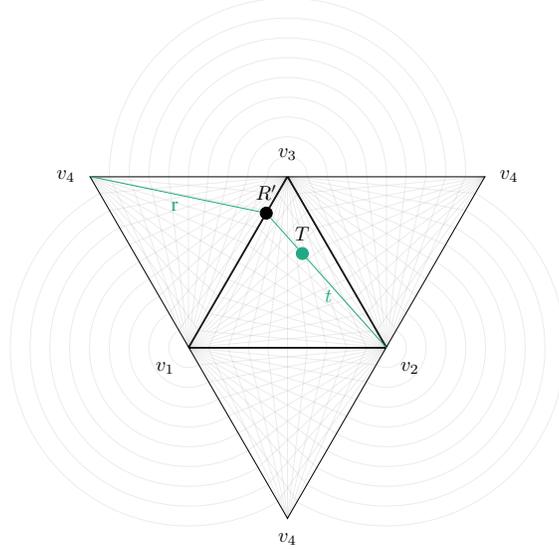

		While the preimage of the map~$\phi$ is clearly the $(J-1)$-simplex~$\Delta$, determining the exact image is not as straightforward.
		First of all, the function~$\phi$ takes a $J$-dimensional point $x$ and performs $J$ perspective projections, yielding a vector
		$$
		    \phi(x) = 
		    \big(\psi_1(x), \ldots, \psi_J(x)\big) \defeq
		    \big(\phi(x)_1, \ldots, \phi(x)_J\big).
		$$
		Each element of this vector is an element of the corresponding facet,  $\psi_j(x)\defeq\phi(x)_j\in\sigma_{-j}$, making the vector of projections an element of the product space of the facets~$\prod\limits_{j=1}^J\sigma_{-j}$.
		However, due to the invariance of the barycentric coordinate ratios to perspective projection, not all combinations in the product space are possible images (cf. \autoref{theo:preimage-unlabeled}).
		Instead, we need to restrict the image to objects where all combinations of projections $\phi(x)_n, \phi(x)_m, (n, m)\in\{1,\ldots,J\}^2$ are compatible to each other according to the $\rel$ relation (see below).
		The image of $\Delta$ under the function~$\phi$ follows as
		\begin{equation}\label{eq:image-domain-simplex projection}
		    \mathrm{Img}_{\phi}(\Delta) = \left\{\big(\phi(x)_1,\ldots,\phi(x)_J)\in\prod\limits_{j=1}^J \sigma_{-j}\,\middle\vert\, \phi(x)_n\rel\phi(x)_m\ \forall (n,m)\in \{1,\ldots,J\}^2 \right\}
		\end{equation}
  
\subsection{Proof of Theorem 1}\label{app:proof-theorem-1}

\begin{proof}
        The proof will show that $\phi$ is a bijective map by proving that $\phi$ is invertible.
        
		Let $x=(\pi_1,\ldots,\pi_J)\in\Delta^{J-1}$ and $\phi(x)$ be the projections of $x$ onto the respective facets as defined above.
		The projection on each facet~$\sigma_j$ is described by the renormalized barycentric coordinates after removing the $j^{\text{th}}$ component
		\begin{equation}
		    \psi_j(x) = (\pit_1, \ldots, \pit_{j-1}, \pit_{j+1}, \ldots, \pit_J)
		\end{equation}
		by definition in \autoref{eq:bijection-Delta-K} and \autoref{eq:perspective-projection}.
		By repeatedly removing another component and re-normalizing the coordinates, we can extract numerical values 
		$$
			\nicefrac{\pi_n}{\pi_m}=r_{n,m}
		$$ 
		for the ratios of all pairs of barycentric coordinates of the original point~$x$ because the barycentric coordinate \emph{ratios} are invariant to projection.
		In the following, we will only consider the ratios of subsequent components, i.e., $\{\nicefrac{\pi_1}{\pi_2},\nicefrac{\pi_2}{\pi_3},\ldots,\nicefrac{\pi_{J-1}}{\pi_J}\}$.
		The other ratios are not required to solve the problem at hand.
		Recall that the constraint $\sum_{j=1}^J\pi_j=1$ still holds for the barycentric coordinates of $x$.
		This yields a system of $J$ equations
		\begin{equation}
		    \myarray{
		    \pi_1 / \pi_2  = r_{1,2}\\
		    \pi_2 / \pi_3  = r_{2,3}\\
		    \hspace*{1cm}\vdots \\
		    \pi_{J-1} / \pi_{J}  = r_{J-1,J}\\
		    \pi_1 + \ldots + \pi_J  = 1
		    }
		    \Rightarrow
		    \myarray{
		    \pi_1 = r_{1,2}\pi_2\\
		      \pi_2 = r_{2,3}\pi_3\\
		      \hspace*{1cm}\vdots\\
		      \pi_{J-1} = r_{J-1,J}\pi_J\\
		      \pi_1+\ldots+\pi_J = 1
		    }
		    \Rightarrow
		    \myarray{
		    \pi_1 - r_{1,2}\pi_2 = 0\\
		    \pi_2 - r_{2,3}\pi_3 = 0\\
		    \hspace*{1cm}\vdots\\
		    \pi_{J-1} - r_{J-1,J}\pi_J = 0\\
		    \pi_1+\ldots+\pi_J = 1
		    }
		\end{equation}
		with $J$ unknowns and the matrix representation
        \begin{equation}
            \begin{aligned}
            \underbrace{
              \begin{pmatrix}
              1 & -r_{1,2} & 0 & \cdots & 0 & 0\\
              0 & 1 & -r_{2,3} & \cdots & 0 & 0\\
              \vdots & \vdots & \vdots & \ddots & \vdots & \vdots \\
              0 & 0 & 0 & \cdots & 1 & -r_{J-1,J}\\
              1 & 1 & 1 & \cdots & 1 & 1
              \end{pmatrix}
             }_{=\mathbf{A}\in\mathbb{R}^{J\times J}}
              \begin{pmatrix}
              \pi_1 \\
              \pi_2 \\
              \vdots \\
              \pi_{J-1}\\
              \pi_J
              \end{pmatrix}
              =
              \begin{pmatrix}
              0 \\
              0 \\
              \vdots\\
              0 \\
              1
              \end{pmatrix}
            \end{aligned}
        \end{equation}
		which can be solved since $\mathbf{A}\in\mathbb{R}^{J\times J}$ clearly has full rank $J$.
		The solution $(\pi_1,\ldots,\pi_J)^{\top}=\mathbf{A}^{-1}(0, \ldots, 0, 1)^{\top}$, in turn uniquely defines $x$ through its barycentric coordinate representation
		$x=(\pi_1,\ldots,\pi_J)=\sum_{j=1}^J\pi_j v_j\in\Delta^{J-1}$.
	
		We conclude the general case for arbitrary $J$ with two remarks.
		First, the barycentric coordinate ratios do not need to be extracted for subsequent components.
		It is sufficient if $J-1$ independent ratios are calculated to determine $J-1$ unknowns, while the last unknown is solved through the constraint $\sum\pi_j=1$.
		What is more, this implies that any unknown can be solved through the sum-to-one constraint, and it does not need to be $\pi_J$.
		After all, the ordering of the components~$1,\ldots,J$ is arbitrary and we can always re-arrange indices to match the notation in the proof above.
		
		Second, it is not necessary to use the projections onto all facets~$\sigma_{-j}$. 
		In the argumentation above, only $J-1$ independent ratios (and the sum-to-one constraint) are required to recover the original point~$x$.
		$J-1$ independent ratios can, in turn, be extracted from the projections on exactly two different facets:
		From the first projection~$\psi_n(x)=(\pit_1, \ldots,\pit_{n-1},\pit_{n+1},\ldots,\pit_J)$ onto the facet~$\sigma_{-n}$, all necessary ratios except a ratio $\nicefrac{\pi_n}{\pi_{m}}, m\neq n$ involving $\pi_n$ can be extracted.
		However, the ``final'' independent ratio including $\pi_n$ to solve for $\pi_n$ can be extracted from the projection~$\psi_m(x)$ onto the other facet $\sigma_{-m}$ if $n\neq m$.
		This means that the projections onto only two facets must always suffice to recover the original point~$x$ regardless of the dimensionality $J$.
		
		\end{proof}

\subsection{Proof of Theorem 2}\label{app:proof-theorem-2}
\paragraph{Rouché-Capelli Theorem}
		One critical argument in the proof builds on a corollary of the Rouch\'{e}-Capelli theorem \autocite{safarevic_linear_2013}, which we will denote in the following:
        ``In an euclidean space~$\mathbb{R}^{J-1}$, $J$ hyperplanes~$Z_1, \ldots, Z_J$ can have zero, one, or infinitely many concurrencies''.
        The $J$ hyperplanes are defined by $Z_j: a_{j,1}x_1 + \ldots + a_{j, J-1}x_{J-1} = c_j$, and the concurrency (intersection of \emph{all} hyperplanes simultaneously) is the solution to the following system of equations:
        \begin{equation}
        \underbrace{
            \begin{pmatrix}
                a_{1,1} & \cdots & a_{1, J-1} \\
                \vdots & \ddots & \vdots \\
                a_{J,1} & \cdots & a_{J, J-1} \\
            \end{pmatrix}
        }_{\mathbf{A}
        \in\mathbb{R}^{J\times (J-1)}
        }
            \begin{pmatrix}
                x_1 \\
                \vdots \\
                x_{J-1}
            \end{pmatrix}
            =
            \underbrace{\begin{pmatrix}
                b_1\\
                \vdots\\
                b_J
            \end{pmatrix}
            }_{b
            \in\mathbb{R}^{J}
            }
        \end{equation}
        
        The Rouch\'{e}-Capelli theorem \autocite{safarevic_linear_2013} states that this system has
        \begin{enumerate}[label=(\roman*)]
            \item no solution, if and only if the rank of its coefficient matrix~$\mathbf{A}$ equals the rank of the augmented matrix~$[\mathbf{A}|b]$;
            \item exactly one solution if and only if $\mathrm{Rank}(\mathbf{A})=J-1$; and
            \item infinitely many solutions otherwise.
        \end{enumerate}

\paragraph{Proof of the Theorem}
    \begin{proof}
        The proof will show that $\Phi$ is a bijective mapping by showing that there is exactly one solution to label the projections onto the facets (up to permutations), and reduce the problem to \autoref{theo:preimage-matching-points}.
        
        \begin{figure}[t]
            \centering
            \begin{subfigure}[t]{0.48\linewidth}
            \includegraphics[width=1.0\linewidth]{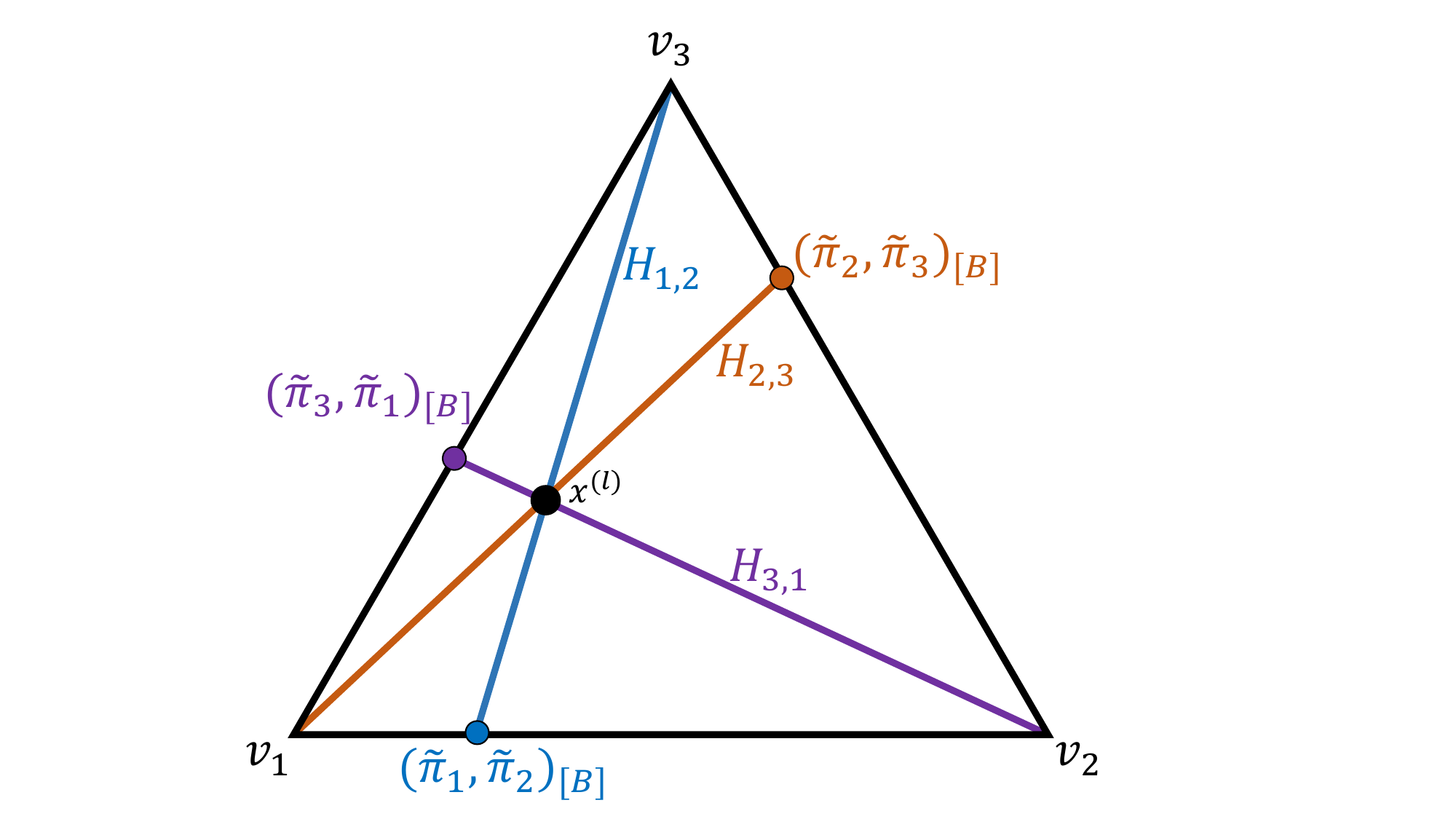}
            \caption{In $\Delta^2$, the hyperplanes which are implied by the ratios on the edges are lines.}
            \label{fig:set-of-points-hyperplane}
            \end{subfigure}
            \hfill
            \begin{subfigure}[t]{0.48\linewidth}
            \includegraphics[width=1.0\linewidth]{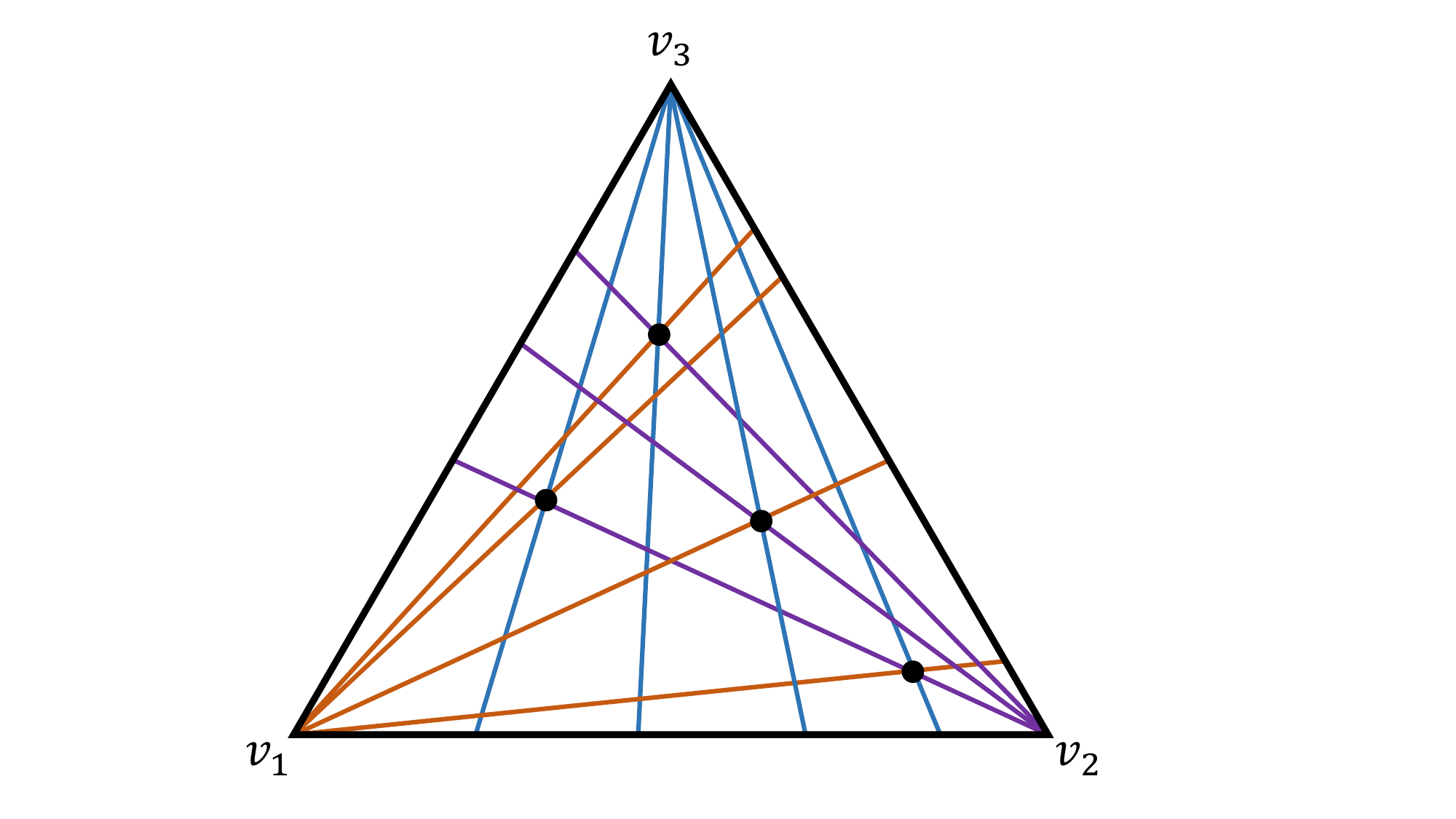}
            \caption{Finding concurrencies of hyperplanes (lines in this case) is equivalent to identifying consistent points in the preimage. Note that hyperplanes originating from the same edge may be equal (bottom-most purple line), but not parallel and unequal.}
            \label{fig:set-of-points-equivalence}
            \end{subfigure}
            \caption{Illustrations for \autoref{theo:preimage-unlabeled}.}
        \end{figure}
        
        As in \autoref{theo:preimage-matching-points}, we can repeatedly remove components from all projections and re-normalize the coordinates to extract numerical values~$\nicefrac{\pi_n}{\pi_m}=r_{n,m}$ for the ratios of all pairs of barycentric coordinates of all projections.
        Each ratio~$r_{n, m}^{(l)}=\nicefrac{\pi_n^{(l)}}{\pi_m^{(l)}}$ implies a hyperplane~$H\in\mathbb{R}^{J-2}$ through the points~$(\pit_n^{(l)}, \pit_m^{(l)})$ and the set of points~$\{v_{a}\}_{a\notin\{n, m\}}$ of all vertices of the $(J-1)$-simplex which have been marginalized out to obtain the ratio~$r_{n, m}^{(l)}$ (see \autoref{fig:set-of-points-hyperplane}). All points~$h\in H\cap\Delta$ in the union of $H$ and the simplex are compatible with the point~$(\pit_n^{(l)}, \pit_m^{(l)})$ on the edge~$\overline{v_nv_m}$: $$\forall h\in H\cap \Delta: h\rel(\pit_n^{(l)}, \pit_m^{(l)}).$$
        
        Finding intersections of hyperplanes in the simplex $\Delta$ is equivalent to finding points which are compatible with the ratios inducing these hyperplanes.
        Consequently, finding a point where hyperplanes induced by $J$ consecutive ratios forming a cycle (i.e., $r_{1,2},\ldots,r_{J-1, J}, r_{J, 1}$) intersect is equivalent to identifying an original point in the preimage which is compatible with all points on the edges with barycentric coordinates of the ratios (see \autoref{fig:set-of-points-equivalence} for an illustration with $J=3$).
        On each linearly independent edge~$\overline{v_1v_2},\ldots,\overline{v_Jv_1}$, there is one extracted ratio (and consequently one implied hyperplane) per original point~$x^{(l)}$.
        Thus, there is a total of $L$ hyperplanes on each linearly independent edge, implied by the set of original points.
        The hyperplane implied by the ratio~$r_{n, m}^{(l)}$ will be referred to as $H_{n,m}^{(l)}$.
        The resulting structure of hyperplanes follows as
        \begin{equation}\label{eq:sets-of-points-hyperplane-definition}
        \underbrace{\{H_{1,2}^{(1)}, \ldots, H_{1,2}^{(L)}\}}_{\text{edge}\;v_1v_2},
        \ldots, 
        \underbrace{\{H_{J-1,J}^{(1)}, \ldots, H_{J-1,J}^{(L)}\}}_{\text{edge}\;v_{J-1}v_J},
        \underbrace{\{H_{J,1}^{(1)}, \ldots, H_{J,1}^{(L)}\}}_{\text{edge}\;v_{J}v_1}
        \end{equation}
        when we group the $L$ hyperplanes implied by a ratio~$r_{n, m}$ in a set, for a total of $J$ sets of $L$ hyperplanes each.
        Finding consistent original points is now equivalent to finding concurrencies of $J$ hyperplanes, each from one of the $J$ sets above, see \autoref{fig:set-of-points-equivalence}.
        What is more, no two $J$ hyperplanes in such an arrangement can be parallel or equal because of the geometry of the simplex.
        However, pairs of hyperplanes within one of the $J$ sets may be equal, yet they cannot be parallel and unequal (see \autoref{fig:set-of-points-equivalence}).
        It remains to be proven that there are exactly $L$ concurrencies where $J$ hyperplanes---one of each of the sets in \autoref{eq:sets-of-points-hyperplane-definition}---intersect.

        \textbf{Statement:} ``There are exactly $L$ concurrencies between the $J$ sets of $L$ hyperplanes each, implied by the ratios of $L$ images according to \autoref{eq:sets-of-points-hyperplane-definition}''.
        We prove this statement by first arguing that there are at least $L$ concurrencies, and then show that there cannot be more than $L$ concurrencies.
        \autoref{fig:set-of-points-equivalence} shows an example where the statement is clearly true. 
        Below, we prove that it holds for all possible cases with arbitrary $L$ and $J$.
        
        \begin{enumerate}[label=(\Roman*)]
        \item \textbf{There are at least $L$ concurrencies.}
        Every preimage point~$x^{(l)}$ has a consistent ratio representation, and thus there must clearly be a concurrency at each point~$x^{(l)}$, resulting in at least $L$ concurrencies.
        \item \textbf{There are not more than $L$ concurrencies.}
        We prove this by contradiction.
        Assume there was an additional $L+1^{\text{st}}$ concurrency~$z\notin\{x^{(l)}\}$.
        This means that $J$ hyperplanes---one originating from each edge---intersect at the point~$z\in\Delta$.
        Call these hyperplanes $Z_1, \ldots, Z_J$ with $Z_j \in \{ H_{j,j+1}^{(l)} \}$ for $j \neq J$ and $Z_J \in \{ H_{J,1}^{(l)} \}$ for $j = J$ as in \autoref{eq:sets-of-points-hyperplane-definition}.
        Each of the $L$ concurrencies~$\{x^{(l)}\}$, which must exist as argued in \textbf{(I)}, has a set of $J$ corresponding hyperplanes which intersect at $x^{(l)}$ by definition.
        Because one hyperplane of each set is induced by each original point, the mapping from an original point (one of the $L$ concurrencies) to $J$ hyperplanes (one of each set of $L$ concurrencies~$\{x^{(l)}\}$ as above) is surjective (exhaustive) on the set of all $J\cdot L$ hyperplanes.%

\begin{figure}[t]
    		\centering
    		\begin{adjustbox}{width=0.3\textwidth}
    			\includegraphics[]{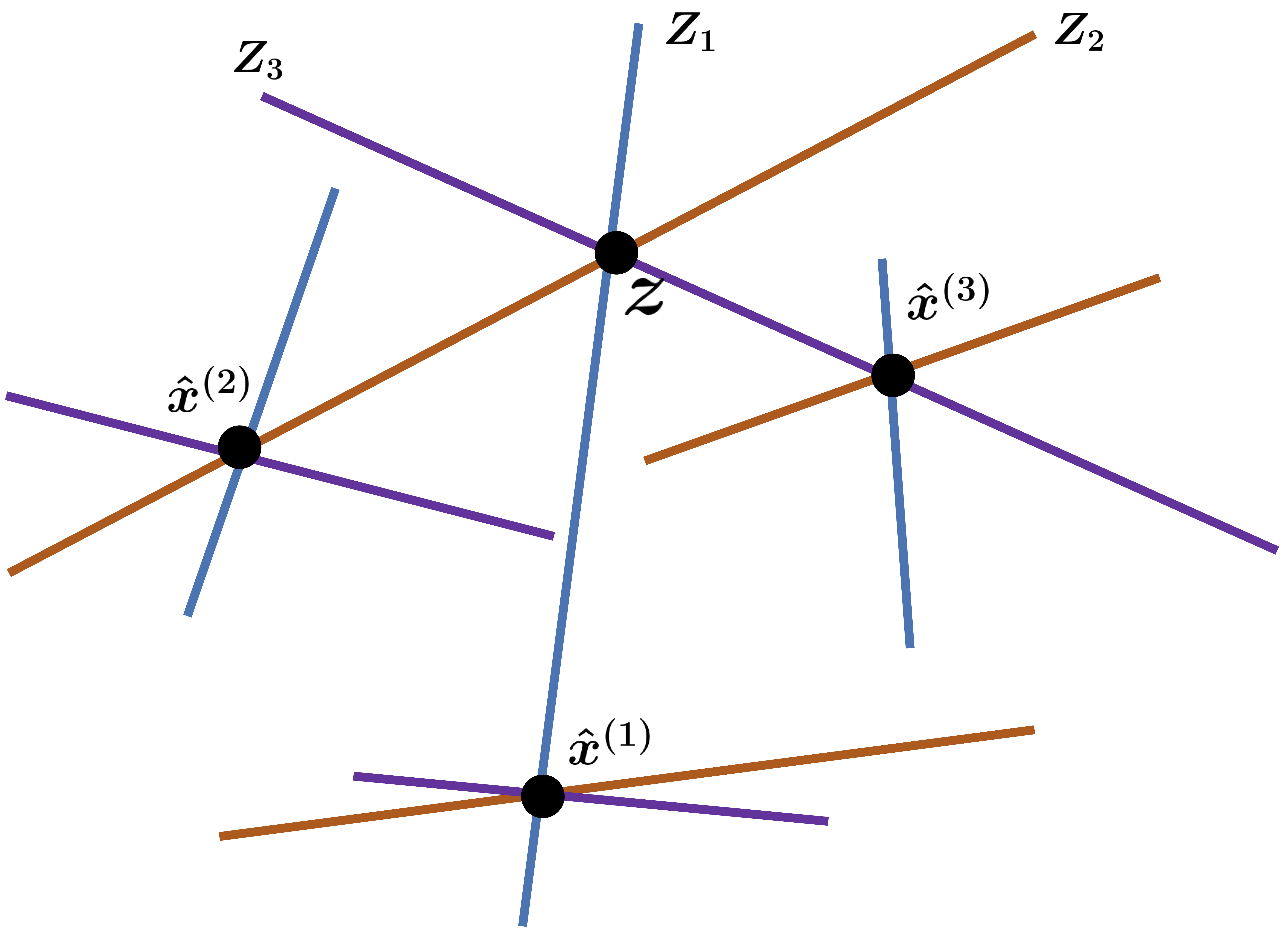}
    		\end{adjustbox}
    	\caption{Illustration of (II) in the proof. Assuming that an additional concurrency $z$ exists, there are corresponding established concurrencies $\hat{x}^{(j)}$, one for each hyperplane (line) $Z_j$, $j = 1, 2, 3$. For each hyperplane (line) $Z_j$, there must be two other hyperplanes (lines) that share established concurrencies $\hat{x}^{(j)}$ with $Z_j$ by construction.}
    	\label{fig:z-and-established}
\end{figure}

        Consequently, each of the hyperplanes~$Z_1, \ldots, Z_J$, which intersect at the ``additional'' concurrency $z$, has a corresponding ``established'' concurrency with $J-1$ other hyperplanes from the set of $L$ concurrencies from above. For example, if $Z_1 = H_{1,2}^{(l')}$, then $x^{(l')} \in \{ x^{(l)} \}$ is an established concurrency corresponding to $Z_1$.
        The phrase ''established`` emphasizes that we already know about these $L$ concurrencies through \textbf{(1)}.
        Now consider these $J$ established concurrencies which correspond to $Z_1, \ldots ,Z_J$, which we call $\hat{x}^{(1)}, \ldots, \hat{x}^{(J)}$, all in $\{x^{(l)}\}$.
        One of the following cases must occur:
        \begin{enumerate}[label=(\alph*)]
            \item \textbf{At least one pair of established concurrencies is equal}, $\exists n,m\in\{1,\ldots,J\}^2, n\neq m: \hat{x}^{(n)} = \hat{x}^{(m)}$. This means that the hyperplanes~$Z_n$ and $Z_m)$ intersect at $\hat{x}^{(n)}=\hat{x}^{(m)}$, and by definition of $z$ they also intersect at $z$.
            The corollary on the Rouch\'{e}-Capelli theorem implies that either $z=\hat{x}^{(n)}=\hat{x}^{(m)}\in\{x^{(l)}\}$ (contradiction; $z$ is assumed to be a concurrency beyond $\{x^{(l)}\}$) or the two hyperplanes~$Z_n$ and $Z_m$ are equal (contradiction).
            \item \textbf{All established concurrencies are different from each other}, $\hat{x}^{(n)}\neq \hat{x}^{(m)}\;\forall n, m\in\{1,\ldots,J\}^2,n\neq m$.
            The fact that $z=(\zeta_1,\ldots,\zeta_J)$ shares a hyperplane with each of the $J$ established concurrencies~$\hat{x}^{(1)}, \ldots, \hat{x}^{(J)}$---each with barycentric coordinate representation~$(\hat{\pi}_1^{(j)}, \ldots, \hat{\pi}_J^{(j)})$---translates to the ratio equalities in barycentric coordinates
            \begin{equation}
            \frac{\zeta_1}{\zeta_2} = \frac{\hat{\pi}^{(1)}_1}{\hat{\pi}^{(1)}_2},\quad
            \frac{\zeta_2}{\zeta_3} = \frac{\hat{\pi}^{(2)}_2}{\hat{\pi}^{(2)}_3},\quad
            \ldots,\quad
            \frac{\zeta_J}{\zeta_1} = \frac{\hat{\pi}^{(J)}_J}{\hat{\pi}^{(J)}_1}.
            \end{equation}
            Generalized Ceva's theorem \autocite{buba-brzozowa_cevas_2000} guarantees that 
            \begin{equation} \frac{\zeta_1}{\zeta_2}\,\frac{\zeta_2}{\zeta_3}\cdots\frac{\zeta_J}{\zeta_1}=1
            \end{equation}
            because $z$ is a concurrency by definition.
            The inverse direction of generalized Ceva's theorem states that the hyperplanes~$Z_1, \ldots, Z_J$ have a concurrency~$\hat{x}$ since the following equality holds:
            \begin{equation}
            \frac{\hat{\pi}^{(1)}_1}{\hat{\pi}^{(1)}_2}\,
            \frac{\hat{\pi}^{(2)}_2}{\hat{\pi}^{(2)}_3}
            \cdots
            \frac{\hat{\pi}^{(J)}_J}{\hat{\pi}^{(J)}_1}=1.
            \end{equation}
            It follows from the corollary of the Rouch\'{e}-Capelli theorem that either $\hat{x}=z$ (contradiction) or that at least two of the hyperplanes are equal (contradiction).
        \end{enumerate}
    \end{enumerate}
\end{proof}

\end{document}